\documentclass[twocolumn]{aastex63}
\shorttitle{Formation of deuterated glycolaldehyde}
\shortauthors{Vazart et al.}
\graphicspath{{./}{figures/}}
\usepackage{appendix}

\begin{document}

\title{Quantum chemical computations of gas-phase glycolaldehyde deuteration and constraints to its formation route}

\correspondingauthor{Cecilia Ceccarelli and Nadia Balucani}
\email{cecilia.ceccarelli@univ-grenoble-alpes.fr,nadia.balucani@unipg.it}

\author[0000-0002-8023-3140]{Fanny Vazart}
\affiliation{Univ. Grenoble Alpes, CNRS, Institut de Planétologie et d’Astrophysique de Grenoble (IPAG), 38100 Grenoble, France}

\author[0000-0001-9664-6292]{Cecilia Ceccarelli}
\affiliation{Univ. Grenoble Alpes, CNRS, Institut de Planétologie et d’Astrophysique de Grenoble (IPAG), 38100 Grenoble, France}

\author[0000-0002-4741-5866]{Dimitrios Skouteris}
\affiliation{Master-Tec, Via Sicilia 41, 06128 Perugia, Italy}

\author[0000-0001-5121-5683]{Nadia Balucani}
\affiliation{Dipartimento di Chimica, Biologia e Biotecnologie, Università degli Studi di Perugia, 06123 Perugia, Italy}

%%%%%%%%%%%%%%%%%%%%%%%%%%%%%%%%%%%%%%%%
%% Mark off the abstract in the ``abstract'' environment. 
\begin{abstract}
Despite the detection of numerous interstellar complex organic molecules (iCOMs) for decades, it is still a matter of debate whether they are synthesized in the gas-phase or on the icy surface of interstellar grains.
In the past, molecular deuteration has been used to constrain the formation paths of small and abundant hydrogenated interstellar species. More recently, the deuteration degree of formamide, one of the most interesting iCOM, has also been explained in the hypothesis that it is formed by the gas-phase reaction NH$_2$ + H$_2$CO.
In this article, we aim at using molecular deuteration to constrain the formation of another iCOM, glycolaldehyde, which is an important prebiotic species.
More specifically, we have performed dedicated electronic structure and kinetic calculations to establish the glycolaldehyde deuteration degree in relation to that of ethanol, which is its possible parent species according to the suggestion of Skouteris et al. (2018).
We found that the abundance ratio of the species containing one D-atom over the all-protium counterpart depends on the produced D isotopomer and varies from 0.9 to 0.5.
These theoretical predictions compare extremely well with the monodeuterated isotopomers of glycolaldehyde and that of ethanol measured towards the Solar-like protostar IRAS 16293-2422, supporting the hypothesis that glycolaldehyde could be produced in the gas-phase for this source.
In addition, the present work confirms that the deuterium fractionation of iCOMs cannot be simply anticipated based on the deuterium fractionation of the parent species but necessitates a specific study, as already shown for the case of formamide.

\end{abstract}

%% Keywords should appear after the \end{abstract} command. 
%% See the online documentation for the full list of available subject
%% keywords and the rules for their use.

%% We recommend that authors also use the natbib \citep
%% and \citet commands to identify citations.  The citations are
%% tied to the reference list via symbolic KEYs. The KEY corresponds
%% to the KEY in the \bibitem in the reference list below. 

%%%%%%%%%%%%%%%%%%%%%%%%%%%%%%%%%%%%%%%%%%%%%%%%%%%%%%%%%%%%%%%%%%%%%%%%%%%%%%%%
%%%%%%%%%%%%%%%%%%%%%%%%%%%%%%%%%%%%%%%%%%%%%%%%%%%%%%%%%%%%%%%%%%%%%%%%%%%%%%%%
%%%%%%%%%%%%%%%%%%%%%%%%%%%%%%%%%%%%%%%%%%%%%%%%%%%%%%%%%%%%%%%%%%%%%%%%%%%%%%%%
\section{Introduction} \label{sec:intro}
Among the more than 270 species that have been detected in the interstellar medium (ISM) so far \citep{McGuire2022-census}, about 40\% are composed of six atoms or more and contain at least a carbon atom.
They are referred  as interstellar Complex Organic Molecules \cite[iCOMs:][]{Herbst2009,Ceccarelli2017}, where the "i" emphasizes that the "complex" adjective only holds in the ISM context. 
iCOMs are significant in prebiotic chemistry research, whose goal is to understand how life could appear on an originally inorganic Earth. 
iCOMs can be considered as precursors or intermediates in the formation of the building blocks of life, such as amino acids or nucleobases for instance. 
The question of their formation in the harsh ISM conditions is, therefore, of paramount importance and far-reaching. 

So far, two major processes responsible for the iCOMs synthesis have been invoked in the literature: reactions occurring on the interstellar icy grain surfaces or in the gas-phase \cite[e.g.][]{Garrod2006,balucani2015,Ceccarelli2022-PP7}.  
Which one dominates and where has been a decades-long quarrel that started soon after the first detection of iCOMs in massive hot cores \citep{Rubin1971,Blake1987} and revived when iCOMs were detected in hot corinos \citep{Cazaux2003,Ceccarelli2004}. 
Traditionally, a major argument to distinguish the gas-phase versus grain-surfaces origin of the gaseous species observed in hot cores/corinos has been the molecular deuteration or deuterium fractionation, namely the abundance ratio between the species containing D-atoms with respect to the same containing H-atoms \cite[e.g.][]{Tielens1983,Charnley1992,Roberts2000a,Roberts2000b,Ceccarelli2001,Ceccarelli2014-PP6}. 
Indeed, since the  D/H ratio at the birth of the Universe is $A_D=2.4\times10^{-5}$ \citep{PlanckCollaboration2016,Cooke2016}, the abundance of D-bearing molecules would be extremely low without a chemistry-triggered deuterium fractionation.
Instead, since the 70s it has been known that many interstellar di- and tri- atomic molecules do show an enhanced abundance of D-bearing molecules with respect to the elemental D/H abundance \cite[e.g.][]{Hollis1976,Penzias1977,Snyder1977}. 
This is (mainly) caused by the dense ISM low temperatures that enhance the H$_2$D$^+$/H$_3^+$ abundance ratio because of the different zero-point energy of the two species: then, all other small deuterated molecules inherit the deuteration from H$_2$D$^+$ \citep{Watson1974,Dalgarno1984}. 

At the beginning of the 2000s, formaldehyde and methanol were the most complex deuterated species detected in the ISM and the detection of doubly deuterated formaldehyde and even triply deuterated methanol \citep{Turner1990,Ceccarelli1998,Parise2003} were used to assess their grain-surfaces origin \cite[e.g.][]{Ceccarelli2001,Parise2006,Taquet2013}.
Indeed, both molecules are (mostly) produced on the grain surfaces thanks to the hydrogenation of frozen CO \cite[e.g.]{Watanabe2002,Watanabe2003,Rimola2014}, so the crucial parameter is the D/H abundance ratio of atoms landing on the grain surfaces \citep{Tielens1983}, which in turn depends on the H$_2$D$^+$/H$_3^+$ abundance ratio in the gas \cite[see e.g. the review by][]{Ceccarelli2014-PP6}.
In summary, until recently, the deuteration of the observed molecules was only connected to the enhanced H$_2$D$^+$/H$_3^+$ in cold gas\footnote{Additionally and/or alternatively, CH$_2$D$^+$ and OD can also enhance molecular deuteration in some species in somewhat warmer gas \cite[$\sim30-100$ K;][]{Roberts2000a,Thi2010}.} which, in turn, is controlled by the so-called thermodynamic isotope effect (TIE, the difference in the zero point energy is such that the isotopic variant H$_3^+$ + HD $\rightarrow$ H$_2$D$^+$ + H$_2$ is slightly exothermic, while the reverse process is slightly endothermic).

The situation has changed in the last years with the detection of D-enriched iCOMs \citep{Coudert2013,Coutens2016,Jorgensen2018,Manigand2019}, because, whatever is their formation route, iCOMs deuteration is not anymore directly connected to the enhanced (gaseous) H$_2$D$^+$/H$_3^+$ abundance ratio but to the deuteration of their parent species. 
In this case, the question is whether the iCOM deuteration is directly inherited from their parent species without any alteration or whether the processes leading from the parent to the daughter species can induce an enrichment or a decrease in the deuteration degree. 

It is worth mentioning here that the rate coefficient of a chemical reaction is expected to change when one of the atoms in the reactants is replaced by one of its isotopes. 
This is the so-called kinetic isotope effect (KIE), not to be confused with the TIE ruling the H$_2$D$^+$/H$_3^+$ abundance. 
A further distinction is made between primary kinetic isotope (in this case, a bond involving the isotopically-labeled atom is being formed or broken) and secondary kinetic isotope effect (when no bond to the isotopically-labeled atom in the reactant is being broken or formed).
Secondary kinetic isotope effects are much smaller than primary kinetic isotope effects.
Finally, and perhaps more importantly, if there is competition between the formation of a D-product or an H-product, it is very important to characterize the reaction mechanism. 
As a matter of fact, if the reaction is an impulsive direct reaction, H-products are favored while, if the reaction is an indirect reaction featuring the formation of a reaction intermediate, since X-D bonds are stronger that X-H bond (where X is an atomic species like carbon or oxygen), D-products are favored.
In other words, to assess the relationship between the deuteration degree of one iCOM and its parent species, the reaction mechanism has to be characterized.

\begin{figure*}[tb]
 \begin{center}
    \includegraphics[scale=0.25]{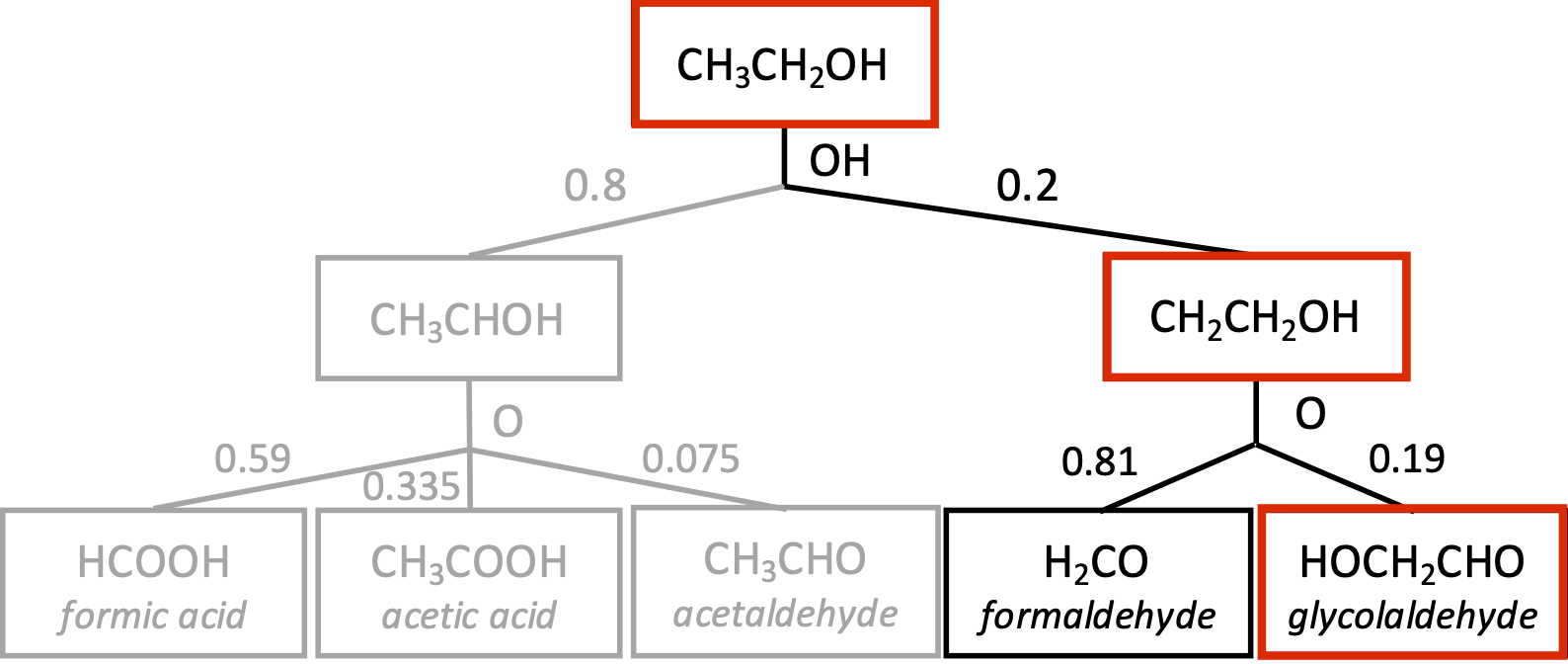}
 \end{center}
    \caption{
    The genealogical tree of ethanol  \citep[adapted from][]{Skouteris2018}
    The numbers in the plot indicate the branching ratios. 
    The present study focuses only on the reactions leading to glycolaldehyde shown in the red boxes.}\label{fig:ethanol-tree}
\label{fig:skouteris2018}
\end{figure*}
For instance, in a recent example, \cite{Skouteris2017} showed that the reaction mechanism of the gas-phase reaction NH$_2$ + H$_2$CO $\rightarrow$ NH$_2$CHO + H is able to explain the measured D/H abundance ratios of formamide (NH$_2$CHO) towards IRAS16293 B. 
This is due to a combination of the D-/H-product branching ratio and KIE.

To the best of our knowledge, the case of formamide is the only case reported in the literature. 
The detection of its D isotopomers was first reported by \cite{Coutens2016} towards the Solar-like protostar IRAS16293-2422 B (hereinafter IRAS16293 B) hot corino.
\cite{Skouteris2017} showed that, if formamide is formed in the gas-phase by the NH$_2$ + H$_2$CO reaction, then:
trans-HCONHD/HCONH$_2 \sim 1/3$ NHD/NH$_2$, cis-HCONHD/HCONH$_2 \sim 1/3$ NHD/NH$_2$ and DCONH$_2$/HCONH$_2  \sim 1/3$ = HDCO/H$_2$C). 
In other words, (i) the singly-deuterated formamide abundance will be a factor three lower than that characterizing the parent species NH$_2$ and H$_2$CO, respectively, while (ii) the relative abundance ratios will reflect those of the parent species (NH$_2$ and H$_2$CO). 
Therefore, the strong message from the Skouteris et al. (2017) study is that the deuterium fractionation of iCOMs from gas-phase reactions depends on the specific characteristics of the formation reaction and cannot be anticipated simply based on the mother species deuteration.
Incidentally, the Coutens et al. observations agree with these theoretical predictions, supporting the case for a gas-phase formation of formamide in the IRAS16293 B hot corino.

In this manuscript, following the \cite{Skouteris2017} study, we report on the case of  glycolaldehyde, the deuterated forms of which have been detected in IRAS16293 B by \cite{Jorgensen2016}. 
As for other iCOMs, two possible formation routes are invoked in the literature: (i) in the gas-phase from gaseous ethanol (CH$_3$CH$_2$OH) following the scheme by \cite{Skouteris2018}, reported in Fig. \ref{fig:ethanol-tree}, or (ii) by the combination of radicals on the grain surfaces as predicted by \cite{Garrod2008} \cite[see also][and references herein]{Simons2020}.
In the present work, we focus on the gas-phase route only, and provide theoretical calculations to predict the deuteration of glycolaldehyde when it is synthesized starting from monodeuterated ethanol.

\begin{figure*}[bt]
    \begin{center}
    \includegraphics[scale=0.25]{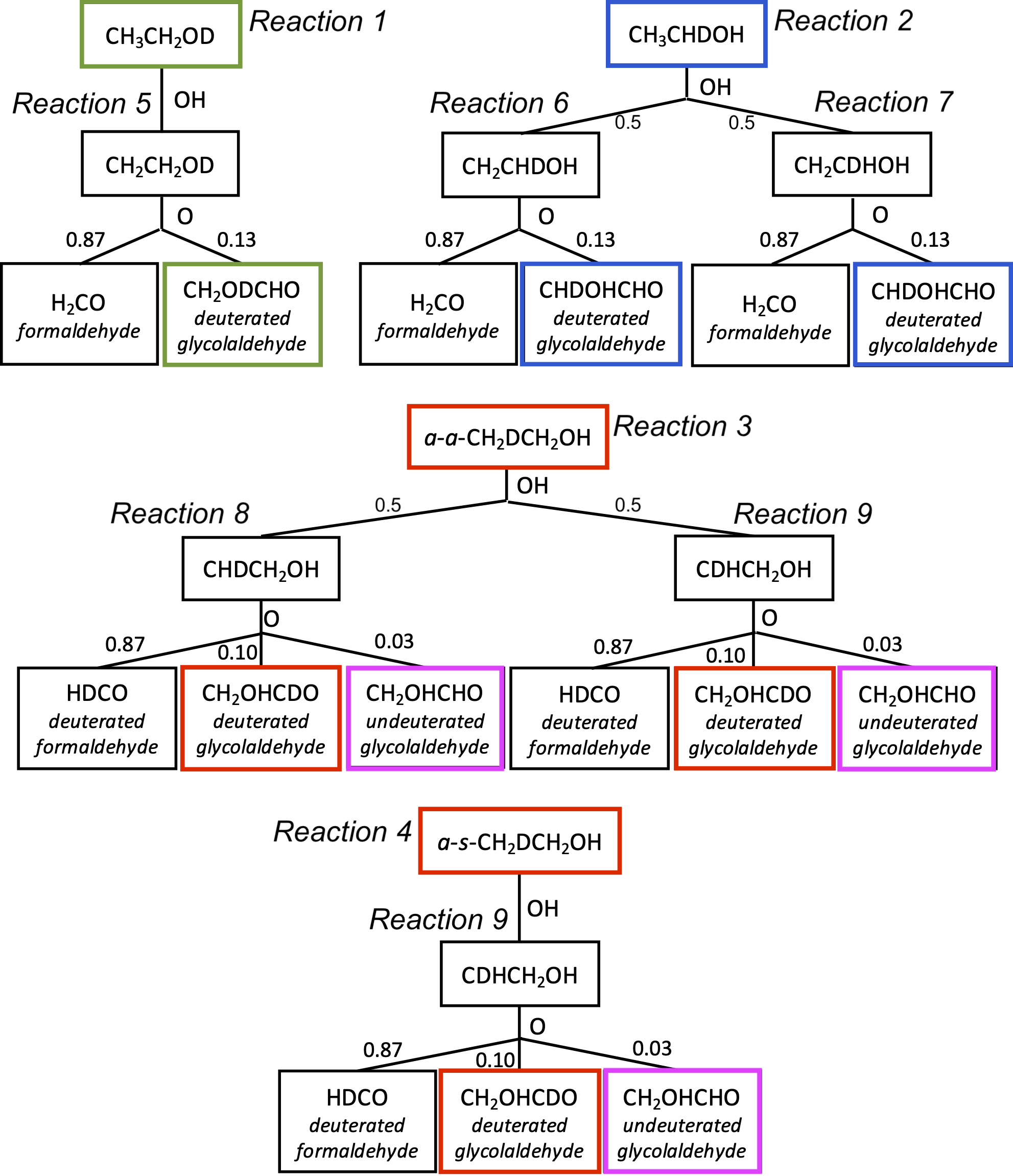}
    \end{center}
    \caption{Summary of the multi-step reactions with their respective major products. 
    Reactions are numbered as in Tab. 1.
    The colored boxes show the parent-daughter relationship.
    The magenta boxes show the channels that lead to undeuterated glycolaldehyde products (8b and 9b).}
    Note that the two types of CH$_2$DCH$_2$OH (a-a and a-s) are characterized by different spectroscopic properties which make them observationally distinguishable. They are characterized, however, by very similar chemical behavior. 
    Branching ratios are also indicated (see text).
    \label{fig:reac-scheme}
\end{figure*}

To this end, we separately follow the effect on the deuteration degree of the two steps shown in Fig. \ref{fig:ethanol-tree}: first, the reactions of partially deuterated ethanol with OH that leads to a combination of five isotopomers/rotamers of CH$_2$CH$_2$OH, and, second, their reactions with atomic oxygen leading to the three deuterated isotopomers of glycolaldehyde. 
While we discuss the consequences of the first step starting from the data available in the literature, we have calculated the rate coefficients and H-/D-product branching ratios associated with the second step in this work.
The studied reactions are listed in Table \ref{tab:reaction-list} and their scheme is shown in Fig. \ref{fig:reac-scheme}.

The article is organised as follows. 
In Section \ref{sec:Review}, we provide some considerations on the first step, namely the hydrogen or deuterium abstraction from deuterated ethanol. We then describe the theoretical methods used for this study in Section \ref{sec:Methods}, and show the results we obtained in Section \ref{sec:Results}. Section \ref{sec:Discussion} is dedicated to a discussion about the astronomical relevance of this study and the comparison between our predictions and observations. Section \ref{sec:Conclusions} concludes the article.

\begin{table*}[tb]
    \centering
    \caption{List of isotopic variants of the OH + CH$_2$CH$_2$OH gas-phase reaction involved in the formation of deuterated of glycolaldehyde following the scheme in \cite{Skouteris2018} also shown in Fig. \ref{fig:skouteris2018}.
    The structural formula of each isotopomer is shown in Fig. 2.
    The reaction numbers follow the sequence of reactions shown in Fig. \ref{fig:reac-scheme}.
    Note that the radical CHDCH$_2$OH can be generated by both reactions 3 and 4 and, therefore, reaction 9 accounts for both. 
    Note that reactions 6 and 7 produce the same isotopomer of glycolaldehyde, as well as reactions 8 and 9.}
    \label{tab:reaction-list}
    \begin{tabular}{llcl}
    \hline
    No. & Reactants & & Products \\
    \hline
    \multicolumn{4}{c}{\textit{Step 1}} \\
    1  &  CH${_3}$CH${_2}$OD + OH   & $\rightarrow$ & CH${_2}$CH${_2}$OD + H$_2$O \\
    2a &  CH${_3}$CHDOH + OH           & $\rightarrow$ & CH${_2}$CHDOH + H$_2$O \\
    2b &                               & $\rightarrow$ & CH${_2}$CDHOH + H$_2$O \\
    3a &  a-a-CH${_2}$DCH${_2}$OH + OH & $\rightarrow$ & CHDCH${_2}$OH + H$_2$O \\
    3b &                               & $\rightarrow$ & CDHCH${_2}$OH + H$_2$O \\
    4  &  a-s-CH${_2}$DCH${_2}$OH + OH & $\rightarrow$ & CHDCH${_2}$OH + H$_2$O \\
    \hline
    \multicolumn{4}{c}{\textit{Step 2}} \\
    5 & CH${_2}$CH${_2}$OD + O & $\rightarrow$ & CH${_2}$ODCHO + H\\
    6 & CH${_2}$CHDOH + O & $\rightarrow$ & CHDOHCHO + H\\
    7 & CH${_2}$CDHOH + O & $\rightarrow$ & CHDOHCHO + H\\
    8a & CHDCH$_2$OH  + O & $\rightarrow$ & CH$_2$OHCDO + H \\
    8b &                  & $\rightarrow$ & CH$_2$OHCHO + D\\
    9a & CDHCH$_2$OH  + O  & $\rightarrow$ & CH$_2$OHCDO + H\\
    9b &                  & $\rightarrow$ & CH$_2$OHCHO + D\\
    \hline
\end{tabular}
\end{table*}
%

%%%%%%%%%%%%%%%%%%%%%%%%%%%%%%%%%%%%%%%%%%%%%%%%%%%%%%%%%%%%%%%%%%%%%%%%%%%%%%%%
%%%%%%%%%%%%%%%%%%%%%%%%%%%%%%%%%%%%%%%%%%%%%%%%%%%%%%%%%%%%%%%%%%%%%%%%%%%%%%%%
%%%%%%%%%%%%%%%%%%%%%%%%%%%%%%%%%%%%%%%%%%%%%%%%%%%%%%%%%%%%%%%%%%%%%%%%%%%%%%%%
\section{The first abstraction step and implications for the deuteration degree} \label{sec:Review}

Following the suggestion by \cite{Skouteris2018}, the first step of the envisaged reaction sequence consists of an H-atom abstraction from ethanol by OH.
The all-protium reaction OH + C$_2$H$_5$OH has been investigated from an experimental and theoretical point of view \citep{Galano2002,Xu2007,Elm2013,Sivaramakrishnan2010}.
The rate coefficients as a function of the temperature (encompassing the range of temperature of interest in the ISM) was determined in two different CRESU (Cinétique de Réaction en Ecoulement Supersonique Uniforme) experiments \citep{caravan2015,Ocana2018}. 
However, this technique is not able to give information on the branching ratios of the different products \citep{cooke2019}.
Theoretical calculations and experimental work have pointed out that both CH$_3$CHOH and CH$_2$CH$_2$OH can be formed \citep{Marinov1999,Carr2011}, while the formation of CH$_3$CH$_2$O has been ruled out \citep{caravan2015}. 
Given the uncertainty on the product branching ratios at low temperatures, \cite{Skouteris2018} have adopted two sets of branching ratios, that is, 0.9/0.7 for CH$_3$CHOH and 0.1/0.3 for CH$_2$CH$_2$OH, respectively (see the two scenarios in Skouteris et al.). 
Here, we have adopted an intermediate value using a yield of 0.8 for CH$_3$CHOH and  of 0.2 for CH$_2$CH$_2$OH.

In this work, we are interested only in the formation of deuterated glycolaldehyde and, therefore, we will consider the part of the ethanol tree that leads to it, \textit{i.e.} the H-abstraction from the methyl group with the formation of the 2-hydroxyethyl radical (the red boxes of Fig. \ref{fig:skouteris2018}). 
To be noted that abstraction reactions are impulsive direct processes.

\begin{figure*}[tb]
    \begin{center}
    \includegraphics[scale=0.18]{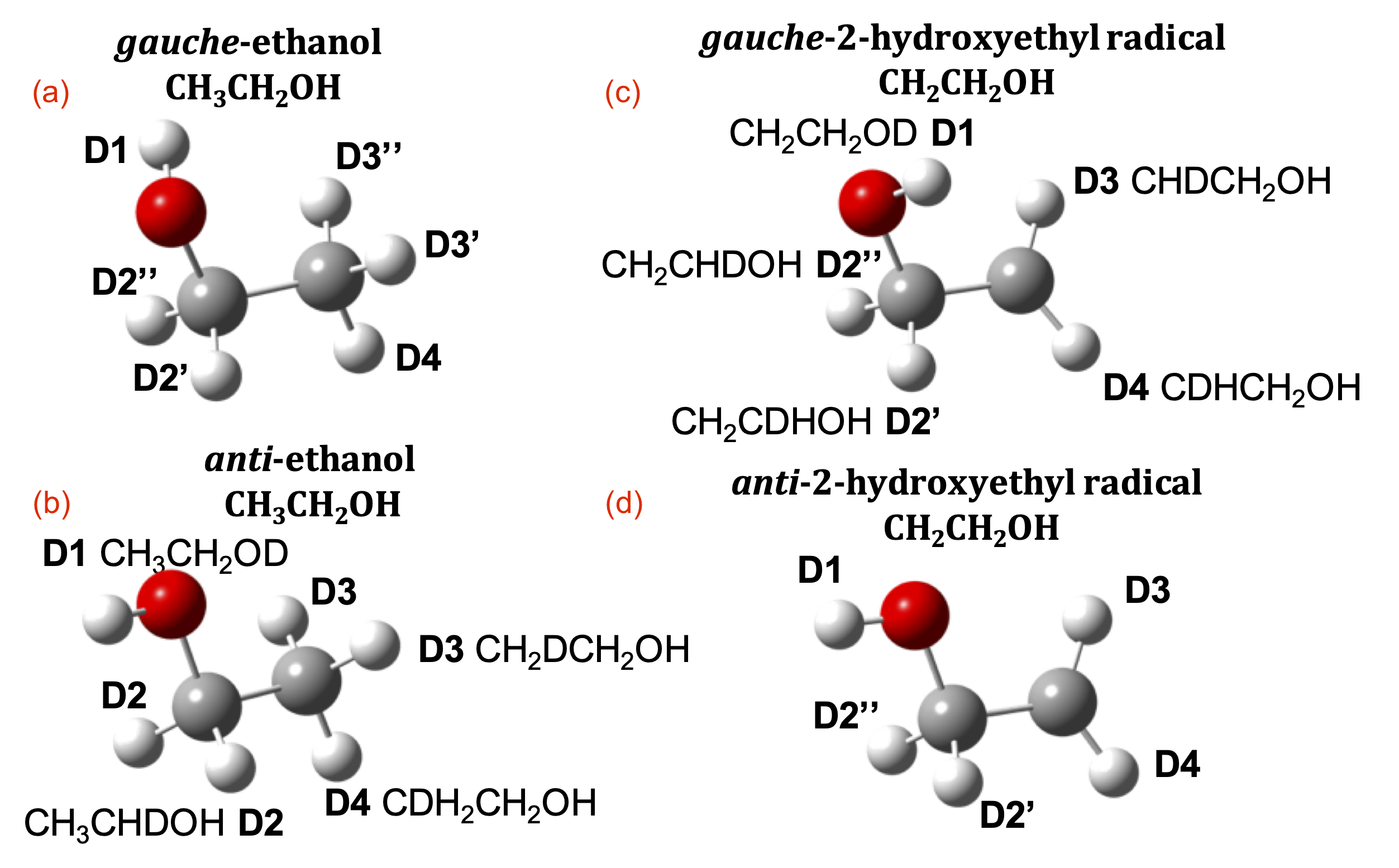}
    \end{center}
    \caption{Different types of deuterated ethanol (left) and deuterated 2-hydroxyethyl radical (CH$_2$CH$_2$OH: right).}
    \label{fig:Eth-radical-deuterated}
\end{figure*}
Ethanol and the 2-hydroxyethyl radical exist under the form of three rotamers regarding the CO bond: the \textit{anti}-, \textit{+gauche}- and \textit{-gauche}- isomers (see Fig. \ref{fig:Eth-radical-deuterated}). 
The two \textit{gauche} isomers are chemically indistinguishable. 
In the case of ethanol, the \textit{anti}- rotamer is the one with the lowest energy content, while in the case of CH$_2$CH$_2$OH the most stable rotamer is the \textit{gauche}-2-hydroxyethyl radical.

If we consider the monodeuterated species of ethanol, because of its four distinct types of hydrogen atoms we have four isotopomers (Fig. \ref{fig:Eth-radical-deuterated}, panels a and b): CH$_3$CH$_2$OD (D1), CH$_3$CHDOH (D2), a-s-CH$_2$DCH$_2$OH (D3) and a-a-CDH$_2$CH$_2$OH (D4).  
Note that a-s-CH$_2$DCH$_2$OH and a-a-CDH$_2$CH$_2$OH have different spectroscopic properties and, therefore, are distinguishable when observed in the ISM. 
However, they are characterized by a very similar chemical behaviour.

Let us now consider the relation between the four isotopomers of ethanol and the isotopomers of the 2-hydroxyethyl radical that can be formed by H- (or D-)abstraction. 
In all cases, we are considering only the abstraction from the methyl group of ethanol because this is the step that leads later to glycolaldehyde. 
Also, since the \textit{gauche}-2-hydroxyethyl radical is the most stable rotamer and the calculations have been done starting from this species (see below), the relation will be shown for this species only. 
Similar reasoning can be applied also to the case of the \textit{anti} rotamers.

If we start from CH$_3$CH$_2$OD, the degree of deuteration remains unchanged in this first step because the D atom is not directly involved in the process and the mechanism is of the direct type (impulsive reaction). 
If compared to the all-protium reaction, the rate coefficient could be slightly affected by secondary KIE, but the effect is expected to be minor.

If we start from CH$_3$CHDOH, the H-abstraction from the terminal methyl group leads to the formation of two isotopomeric radicals (Fig. \ref{fig:Eth-radical-deuterated}, panel c): CH$_2$CHDOH (D2') and CH$_2$CDHOH (D2''). 
In this case, there is no direct competition between the H- or D- abstraction and the deuteration degree is not going to be affected. 
Given the chemical equivalence of the D2 position, we assign a branching ratio of 0.1 to both CH$_2$CHDOH and CH$_2$CDHOH. 
Also in this case, secondary KIE are expected to be negligible.

Finally, in the case of the CH$_2$DCH$_2$OH isotopomers, there is a competition between D-abstraction and H-abstraction. 
However, the H/D-abstraction in the reaction OH + ethanol is characterized by a significant entrance barrier and, while the H-abstraction process is made possible at low temperature by quantum tunneling through that barrier \citep{Ocana2018,caravan2015}, the same is not valid for the D-abstraction. 
The tunneling probability shows exponential dependency on the mass of the particle and doubling the mass of a tunneling H atom by replacing it with deuterium drastically reduces the rate. 
Therefore, only the H-atom will be abstracted by OH since the same process involving D will not benefit from the tunneling effect (primary KIE). 
In this case, as well, the original deuteration degree of ethanol will be retained in the 2-hydroxyethyl radical.

In conclusion, we can say that, for different reasons, the first step does not alter the deuteration degree or the position of the deuterium atom in the 2-hydroxyethyl radical with respect to its parent ethanol molecule. 
Secondary KIE are expected to be negligible in all cases and we can assume that reactions (1-4) are characterized by the same rate coefficients as the all-protium reaction.

After the first step, therefore, we have five possible monodeuterated species of 2-hydroxyethyl radical (there is no symmetry in this case): CH$_2$CH$_2$OD (D1), CH$_2$CHDOH (D2'), CH$_2$CDHOH (D2''), CHDCH$_2$OH (D3), CDHCH$_2$OH (D4).

In the ISM, only the \textit{anti} ethanol rotamer has been observed so far, but the \textit{gauche}-ethanol is expected to be more abundant because of its degeneracy and little energy difference. 
Considering also that the \textit{gauche}-2-hydroxyethyl radical is the more stable rotamer, our theoretical study was performed starting from it. 
This is also motivated by the fact that the H-abstraction is an exothermic reaction that will liberate a large excess energy allowing all possible rotations to occur. 
In any case, we have verified that, if we consider \textit{anti}-2-hydroxyethyl radical as a starting reactant, the global outcome is not changing (see Fig. \ref{fig:PES-ch2ch2oh-o}). 
The addition of atomic oxygen to \textit{anti}-2-hydroxyethyl radical leads indeed to the addition intermediate RI2 (RI stand for Reaction Intermediate) rather than to the addition intermediate RI1, but RI1 and RI2 easily interconvert through the transition state TS1 (TS stands for Transition State) and are expected to be in equilibrium. 
Therefore, no changes are expected whether we start from the \textit{anti} or the \textit{gauche} radical.

%The summary of all the parent and major daughter species of interest in this study is given in Fig. \ref{fig:reac-scheme}. 
%All four types of deuterated ethanol are shown, together with their major products. 

%%%%%%%%%%%%%%%%%%%%%%%%%%%%%%%%%%%%%%%%%%%%%%%%%%%%%%%%%%%%%%%%%%%%%%%%%%%%%%%%
%%%%%%%%%%%%%%%%%%%%%%%%%%%%%%%%%%%%%%%%%%%%%%%%%%%%%%%%%%%%%%%%%%%%%%%%%%%%%%%%
%%%%%%%%%%%%%%%%%%%%%%%%%%%%%%%%%%%%%%%%%%%%%%%%%%%%%%%%%%%%%%%%%%%%%%%%%%%%%%%%
\section{Methods} \label{sec:Methods}

All the computations were carried out using the Gaussian16 suite of programs \citep{g16}. 
Geometry optimizations were performed for every involved species employing the B2PLYP double hybrid functional \citep{Grimme2006}, in conjunction to the aug-cc-pVTZ triple-$\zeta$ basis set \citep{Kendall1992,Woon1993}. 
Semi-empirical dispersion effects were included by means of the D3BJ model of Grimme \citep{Grimme2011}, leading to the so-called B2PLYP-D3/aug-cc-pVTZ level of theory. These optimizations were followed by harmonic vibrational calculations adopting the same method, in order to verify that all intermediates were true minima of the potential energy surface (PES) and that all transition states exhibited a single imaginary frequency. The electronic energies were then reevaluated via the coupled-cluster singles and doubles approximation augmented by a perturbative treatment of triple excitations \cite[CCSD(T),][]{Raghavachari1989} in conjunction to the same basis set. 
This composite method will be hereafter named CCSD(T)/aug-cc-pVTZ//B2PLYP-D3/aug-cc-pVTZ.

As in previous works \citep{Balucani2012,Leonori2013,Skouteris2015,Vazart2015,Skouteris2018}, a combination of capture theory and the Rice-Ramsperger-Kassel-Marcus (RRKM) calculations was used to determine the relevant rate coefficients and branching ratios regarding the reactions following the oxygen addition to the radicals. 
We have used capture theory for the addition of the O($^3$P) atom to the various forms of deuterated hydroxyethyl radicals, whereas for the subsequent steps (isomerizations or bond fissions) energy-dependent rate constants were calculated using the RRKM scheme.
We used an in-house code to solve the master equation in the zero pressure limit.
The quantum tunneling effects were computed following the symmetric Eckart potential treatment, which provides results that do not deviate much from the more accurate instanton method \citep[e.g.][]{Senevirathne_2017_Hdiffusion}.

Given the critical role of the long range interaction in the capture model, we decided to employ a higher accuracy theoretical method based on CASSCF and CASPT2 calculations (OpenMolcas 18.09).
However, we met with considerable difficulty to identify the van der Waals minima in the entrance valley and possible transition states from the van der Waals complex(es) to the first covalently bound intermediate.
Therefore, as a guide to locate the position of the van der Waals energy minimum we resorted to a semiempirical method developed by \cite{pirani2008}, as recently done for other systems by \cite{dearagao2021} and \cite{Marchione2022}. 
This treatment of the entrance channel leads to some variation in the rate coefficient values with respect to those reported in \cite{Skouteris2018}. 
Since the focus of this manuscript is on the deuteration degree of the reaction products, which is not affected by the long-range potential, the details of these calculations and the effect on the global rate coefficients will be the subject of a future publication.

Finally, to obtain the product branching fractions, the master equation was solved at all relevant energies for all systems (to consider the overall reaction scheme), Boltzmann averaging was carried out to obtain temperature-dependent values.

%%%%%%%%%%%%%%%%%%%%%%%%%%%%%%%%%%%%%%%%%%%%%%%%%%%%%%%%%%%%%%%%%%%%%%%%%
%%%%%%%%%%%%%%%%%%%%%%%%%%%%%%%%%%%%%%%%%%%%%%%%%%%%%%%%%%%%%%%%%%%%%%%%%
%%%%%%%%%%%%%%%%%%%%%%%%%%%%%%%%%%%%%%%%%%%%%%%%%%%%%%%%%%%%%%%%%%%%%%%%%
\section{Results} \label{sec:Results}

\subsection{Electronic structure: Potential Energy Surfaces (PESs)} \label{subsec:Electronic structure and vibrational evaluation: Potential Energy Surface (PES)}

\begin{figure*}[t]
    \begin{center}
    \includegraphics[scale=0.5]{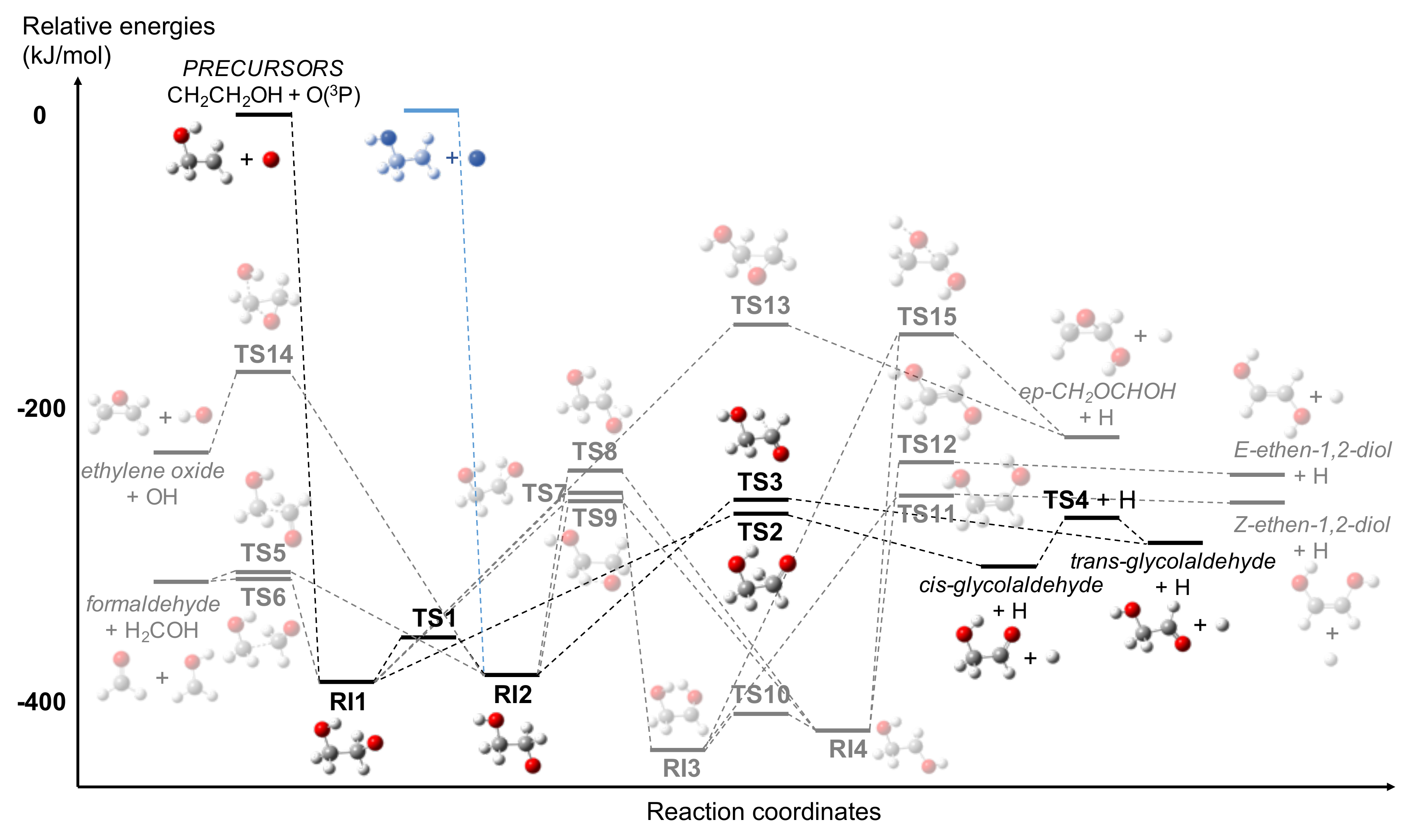}
    \end{center}
    \caption{Proposed full reaction path, starting from CH$_2$CH$_2$OH + O($^3$P), at the CCSD(T)/aug-cc-pVTZ//B2PLYP-D3/aug-cc-pVTZ level of theory. 
    The relative energies are the electronic ones, as they remain equal for every deuterated and undeuterated reactions. RI = reaction intermediate; TS = transition state.
    The detailed energies are given in the Appendix.}
\label{fig:PES-ch2ch2oh-o}
\end{figure*}

%%%%%

Figure \ref{fig:PES-ch2ch2oh-o} shows the minimum energy path following the addition of atomic oxygen O($^3$P) on the 2-hydroxyethyl radical CH$_2$CH$_2$OH, already proposed by \cite{Skouteris2018}. 
All the energies and the coordinates of the involved species are reported in the Appendix. 

As both fragments approach each other, the barrierless addition of oxygen on the \textit{gauche}-2-hydroxyethyl radical leads to the \textbf{RI1}, which is \textit{ca.} 370 kJ/mol more stable than the reactants. Its trans counterpart, the slightly less stable (by 5 kJ/mol) compounds \textbf{RI2}, can easily be reached from the cis, through a transition state (\textbf{TS1}) reflecting a 15 kJ/mol barrier. It is also possible to start from the \textit{anti}- conformer of the radical (\textit{ca.} 2 kJ/mol less stable than the \textit{gauche}) that would lead to \textbf{RI2}.
Both \textbf{RI1} and \textbf{RI2} species are then able to undergo a dissociation into formaldehyde and the CH$_2$OH radical, through the transition states \textbf{TS5} and \textbf{TS6}. 
These dissociation channels exhibit barriers of around 50 kJ/mol. 
Other dissociations can also be observed from both \textbf{RI1} and \textbf{RI2}, leading this time to cis- or trans-glycolaldehyde and H, through \textbf{TS2} and \textbf{TS3}, respectively, that are about 95 kJ/mol higher in energy than their corresponding reaction intermediates. 

Cis- and trans-glycolaldehyde are connected by \textbf{TS4} and require a 30 kJ/mol energy to go from one to the other. 
The cis conformer is the marginally most stable one, by 13 kJ/mol. 
Two epoxidations can also be envisaged. 
The first one, starting from \textbf{RI1}, leads to the epoxide CH$_2$OCHOH + H and exhibits a barrier (represented by \textbf{TS13}) of \textit{ca.} 220 kJ/mol. 
The second one starts from \textbf{RI2}, leads to ethylene oxide and OH and has to go through a \textit{ca.} 200 kJ/mol barrier, represented by \textbf{TS14}. 
Again starting from both \textbf{RI1} and \textbf{RI2} species, one can observe hydrogen migrations from carbon atoms to the oxygen atom bearing the lone electron. 
Both \textbf{TS7} and \textbf{TS8} correspond to hydrogen migrations from the carbon atom linked to the oxygen atom bearing the lone electron and are around 120 kJ/mol more energetic than their corresponding intermediates. 

The \textbf{TS9} transition state can be reached from both \textbf{RI1} and \textbf{RI2} and consists in a hydrogen migration from the carbon that is not linked to the oxygen atom bearing the lone electron. 
It is \textit{ca.} 115 kJ/mol energetically higher than \textbf{RI1} and \textbf{RI2}. \textbf{TS7} and \textbf{TS8} lead to \textbf{RI3} and \textbf{RI4}, respectively, while \textbf{TS9} leads to \textbf{RI4}. 
These intermediates are found to be more than 400 kJ/mol more stable than the precursors and can be linked to each other thanks to the \textbf{TS10} transition state with a small barrier of around 20 kJ/mol. 
Starting from both \textbf{RI3} and \textbf{RI4} compounds, an epoxidation leading to the epoxide CH$_2$OCHOH and H can be envisaged and exhibits \textit{ca.} 250 kJ/mol barriers (\textbf{TS15}). 
The other possible reaction is a dissociation from both \textbf{RI3} and \textbf{RI4} into \textit{Z}- and \textit{E}-ethene-1,2-diol isomers and H, respectively. 
The step from \textbf{RI3} to \textit{Z}-ethene-1,2-diol + H and the one from \textbf{RI4} to \textit{E}-ethene-1,2-diol + H exhibit barriers of about 150 kJ/mol (through the \textbf{TS11} and \textbf{TS12} transition states, respectively), both isomers being found around 260 kJ/mol more stable than the precursors and the \textit{Z} one slightly more stable (by 20 kJ/mol) than its \textit{E} counterpart. 

If we look more carefully at the products that can be obtained via this path, one can see that the most stable ones are formaldehyde + H$_2$COH. The following ones, in decreasing order of stability, are glycolaldehyde + H, \textit{Z}- and \textit{E}-ethene-1,2-diol + H, ethylene oxide + OH and CH$_2$OCHOH + H.

\begin{figure*}[tb]
    \begin{center}
    \includegraphics[scale=0.3]{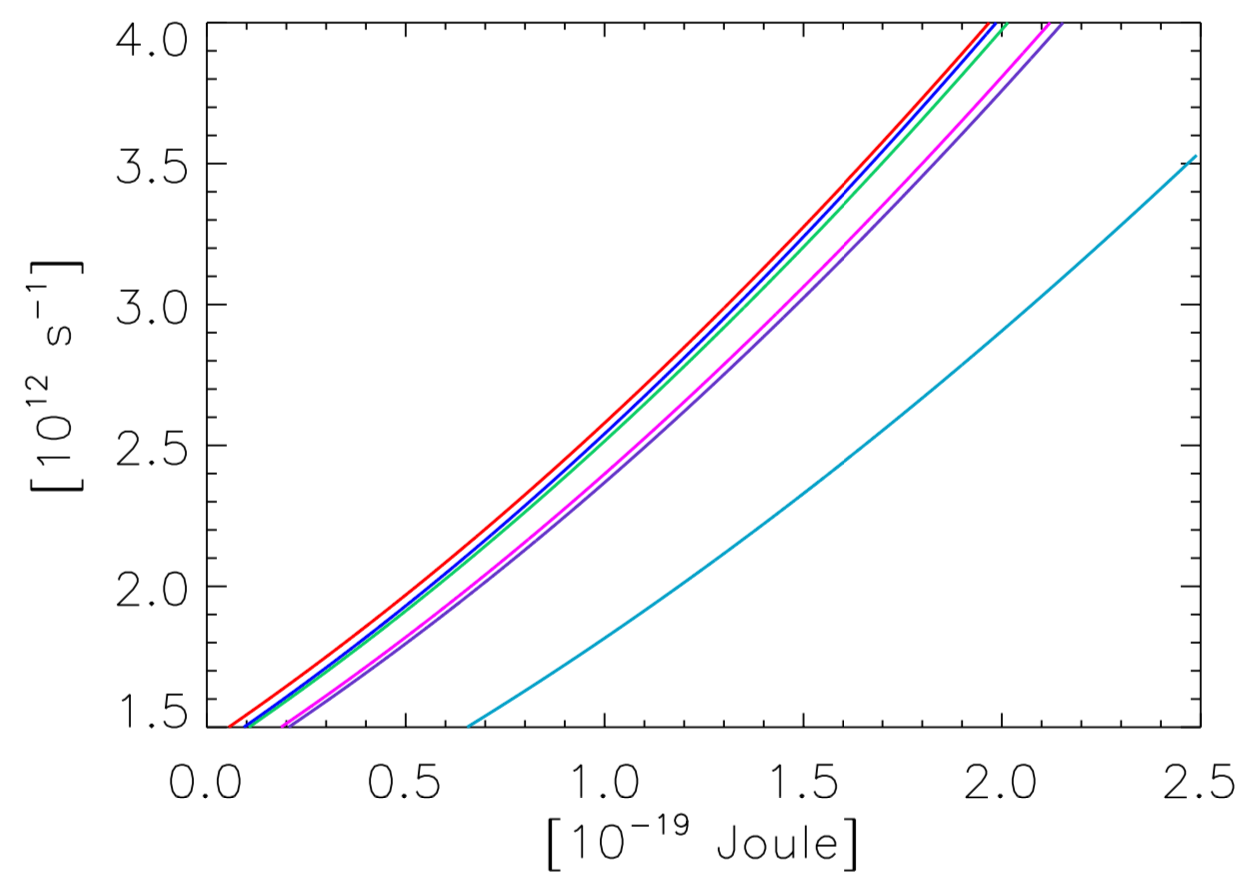}
    \end{center}
    \caption{Unimolecular rate coefficients from RI1 to $cis$-glycoaldehyde via TS2. Red: all-protium reaction. Green: CH$_2$ODCHO from reaction 5. Blue: CHDOHCHO from reaction 6. Magenta: CHDOHCHO from reaction 7. Light blue: CH$_2$OHCDO from reaction 8a. Purple: CH$_2$OHCDO from reaction 9a.}
    \label{fig:unimolecular-rates}
\end{figure*}
When the deuteration of a species is considered, only Zero-Point Energies (ZPEs) are affected as the electronic energy remains identical. 
Indeed, isotopomers carry the same amount of electron(s) and this is the reason why the reaction path of Fig. \ref{fig:PES-ch2ch2oh-o} is valid for all isotopic variants, since the electronic energies remain identical for all five reactions that involve one of the monodeuterated 2-hydroxyethyl radicals.
The energetic distinction between the five paths is then made thanks to zero-point corrections originating from frequencies computations that take into account the masses of the atoms. 
These differences of zero-point corrected energies are reported in Appendix for each case.

The only difference originates from to the presence of a loop between \textbf{RI1}/\textbf{RI2} and \textbf{RI3}/\textbf{RI4} through \textbf{TS7}, \textbf{TS8} and \textbf{TS9}. 
Indeed, for each stage of a reaction, one has to consider the backward and forward steps and if in the all-protium reaction \textbf{TS9} and \textbf{TS7}/\textbf{TS8} link the same intermediates, it is not the case anymore when deuterated species are involved. 
An example of this loop regarding reaction (1) is described in Appendix, together with figures representing all the cases, and were all taken into account for the kinetics calculations. The deuteration propagation between hydroxyethyl radical and glycolaldehyde in every case is also given in Appendix.
We have verified that these loops are not affecting the final outcome, as the dominant channels are those leading from RI1 and RI2 to cis- and trans-glycolaldehyde.

%%%%%%%%%%%%%%%%%%%%%%%
\subsection{Kinetics} \label{subsec:Kinetics}

\begin{figure*}[bt]
    \begin{center}
    \includegraphics[scale=0.2]{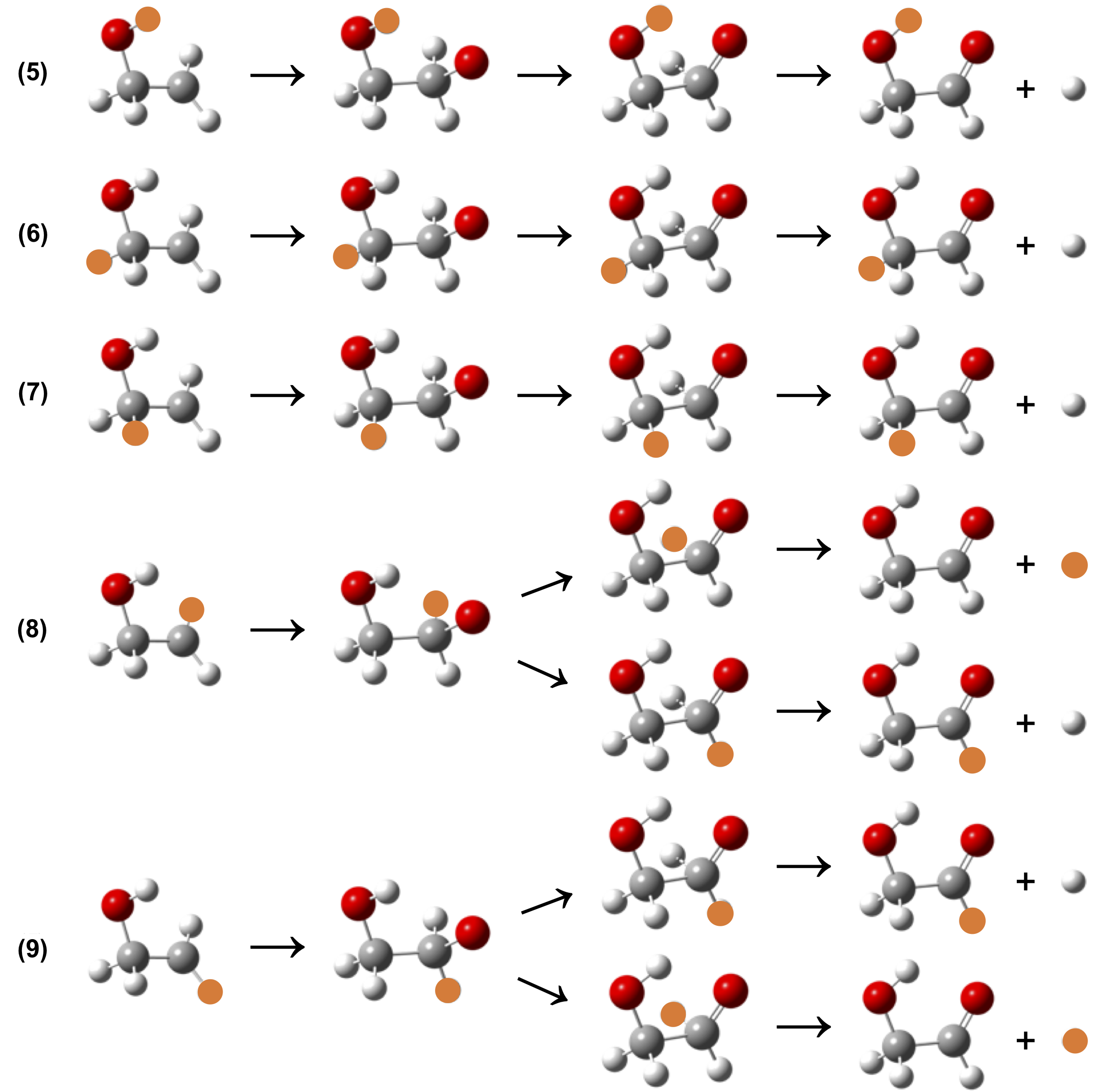}
    \end{center}
    \caption{The structures of the various isotopomers of 2-hydroxylethyl radical, of the relative addition intermediate RI1, of the TS2 connecting RI1 and the products, of the resulting isotopomers of glycolaldehyde for the reactions (5)(9). 
    The D atoms are marked orange.} 
    \label{fig:deuteration-propagation}
\end{figure*}

The main goal of these new calculations is to evaluate the variation of the rate coefficients due to KIE and the D-product branching ratios.
In Fig. \ref{fig:unimolecular-rates}, we have reported the decomposition rate coefficients of the different monodeuterated RI1 into monodeuterated glycolaldehyde (+H) for the reactions (5)-(7), (8a) and 9(a). 
A first aspect to be noted is that the decomposition rate coefficient of RI1 is not very different when the all-protium reaction is compared to the reaction (5) and (6). 
A rather small difference is visible in the case of reaction (7) and (9a) while a more pronounced difference is visible for the reaction (8a). 
An inspection of Fig. \ref{fig:deuteration-propagation} can help to clarify this trend.
Reactions (5)-(7) can only form one isotopomer of glycolaldehyde without competition. 
Furthermore, the C-H bond which is going to be broken does not involve the deuterium atom. 
The similarity with the rate coefficient of the all-protium reaction clearly indicates a very minor contribution from secondary KIE as expected, with the exception of reaction (7) where a small effect can be noted.

The situation is different for the reactions (8) and (9) as in this case we expect competition between the fission of a C-H bond and that of a C-D bond. 
As expected for an indirect reaction, this competition favors the C-H bond breaking over that of C-D. 
Having said that, however, there is a significant difference between the case of reaction (8) and (9). 
In the case of reaction (9), the fission of the C-H bond is so favored that the rate coefficient of the (9a) channel is almost the same as the other rate coefficients of Fig. \ref{fig:unimolecular-rates}. 
Instead, the rate coefficient for reaction (8a) is significantly smaller. 
An inspection of the characteristics of the transition states reported in Fig. \ref{fig:deuteration-propagation} can help in understanding the reason: in order for the H atom to reach the appropriate geometry of the TS2-8a, a significant amount of internal rotation must also occur simultaneously with the bond breaking, increasing the height of the barrier and the rate of its surmounting. 
As a consequence, the yield of the undeuterated glycolaldehyde + D, is significant for reaction (8). 
In the case of reaction (9), instead, the opposite is true because it is the mechanism leading to undeuterated glycolaldehyde + D that features a reorientation of the RI1 intermediate to reach the critical configuration of TS2-9a. 
Indeed, the yield of HOCH$_2$CHO + D is much smaller than in the case of reaction (8). 

In conclusion, the effect of secondary KIE is small in all the cases considered, while primary KIE affecting reaction (8) and (9) has a rather different outcome depending on the details of the potential energy surface and, more specifically in this case, on the difference in the geometry of the intermediates with respect to that of the transition states connecting the intermediates to the products.

In order to be able to compare our new computations with astronomical observations, it is necessary to figure out the amount of every type of deuterated glycolaldehyde that is formed starting from each radical of deuterated ethanol. 
Therefore, we have considered the effect of the first step and multiplied the global rates obtained from each reaction (5-9) by their respective branching ratios resulting from the first step, according to the scheme depicted in Fig. \ref{fig:reac-scheme}. 
In the cases where the same isotopomer of glycolaldehyde has more than one formation reaction, the contributions have been summed up.

%%%%%%%%%%%%%%%%%%%%%%%%%%%%%%%%%%%%%%%%%%%%%%%%%%%%%%%%%%%%%%%%%%%%%%%%%
%%%%%%%%%%%%%%%%%%%%%%%%%%%%%%%%%%%%%%%%%%%%%%%%%%%%%%%%%%%%%%%%%%%%%%%%%
%%%%%%%%%%%%%%%%%%%%%%%%%%%%%%%%%%%%%%%%%%%%%%%%%%%%%%%%%%%%%%%%%%%%%%%%%
\section{Discussion} \label{sec:Discussion}

\subsection{How to compare the theoretical values with astronomical observations}

In this section, we aim to provide guidelines to compare the results of our new computations with  astronomical observations.
As mentioned in Section \ref{sec:Review}, ethanol exists under the form of 3 rotamers: the \textit{anti}-, the \textit{+gauche}- and \textit{-gauche}-. 
However, in the ISM, so far the only detected deuterated conformer is the \textit{anti-} one.
To derive the total abundance of deuterated ethanol (i.e. including the \textit{gauche} rotamers), \cite{Jorgensen2018} suggested to apply a factor of 2.69 to the measured \textit{anti} rotamers abundance \citep[see also][]{manigand2020}.
This practically means that having (so far) only the abundance of the \textit{anti} deuterated conformers, it is necessary to multiply by 2.69 to account for the entire abundance of monodeuterated ethanol. 
Indeed, since we want to compare the abundances of  monodeuterated glycolaldehyde and monodeuterated ethanol, we have to keep in mind that all three starting rotamers of ethanol can lead to glycolaldehyde. 
If a total abundance of the various ethanol isotopologues is ever measured, the factor is not needed anymore.

Given the very close energies of the rotamers of both ethanol and 2-hydroxyethyl radical, we can say that the proposed chemical schemes hold for all rotamers. 
In particular, the very low energy of TS1 warrants an equilibrium population between RI1 and RI2, which are the intermediates formed by either the addition of the O atom to the \textit{gauche-} or \textit{anti-} rotamers of 2-hydroxyethyl radical. 
To address the astrophysical implications, the rate coefficients obtained for reactions (5)-(9) will be divided by the rate coefficient derived for the all-protium reaction. 
Any correction factor necessary to recover the global abundance from the anti rotamer is therefore simplified. 

%%%%%%%%%%%%%%%%%%%%%%%%%%%%%%%%%%%%%%%%%%
\subsection{Application to IRAS16293 B} \label{subsec:discussion-iras16293}

As mentioned in the Introduction, molecular deuteration has proven to be an efficient way to discriminate among the possible different routes of formation of interstellar molecules.
Until very recently, this could only be done for small ($\leq 5$ atoms) molecules  and methanol, in which the molecular deuteration is basically set by the H$_2$D$^+$/H$_3^+$ abundance ratio, which, in turn, is due to thermodynamic isotope effect, TIE.
In large molecules, such as the iCOMs, instead, the deuteration is governed by the kinetic isotope effect, KIE.

The new calculations reported in the present work allow us to test whether the observed deuteration of glycolaldehyde is compatible with its formation in the gas phase from ethanol, following the chain of reactions suggested by \cite{Skouteris2018} and schematically shown in Fig. \ref{fig:ethanol-tree}.
This is also possible thanks to the very sensitive ALMA observations towards the Solar-like protostar IRAS16293 B hot corino obtained by the PILS project \citep{Jorgensen2016}, which detected various isotopomers of deuterated ethanol and glycolaldehyde \citep{Jorgensen2016,Jorgensen2018}.

\begin{deluxetable*}{cccccc}[tb]
    \label{tab:astro-comparison}
    \tablecaption{Comparison of the computed D/H ratios with respect to the observations towards IRAS 16293-2422 by \cite{Jorgensen2016} (glycolaldehyde) and \cite{Jorgensen2018} (ethanol).
    The first two columns report the observed ratio of the measured column densities of the three D-bearing isotopomers of ethanol (D-ethanol) with respect to the H-bearing ethanol (CH$_3$CH$_2$OH). 
    Columns 3 and 4 report the observed ratio of the measured column densities of the three D-bearing isotopomers of glycolaldehyde (D-glycolaldehyde) with respect to the H-bearing glycolaldehyde (CH$_2$OHCHO). 
    Following the scheme of Fig. \ref{fig:reac-scheme}, we then compared the observed and predicted ratios of (CH$_2$ODCHO/CH$_2$OHCHO)/(CH$_3$CH$_2$OD/CH$_3$CH$_2$OH), (CHDOHCHO/CH$_2$OHCHO)/(CH$_3$CHDOH/CH$_3$CH$_2$OH) and (CH$_2$OHCDO/CH$_2$OHCHO)/(CH$_2$DCH$_2$OH/ CH$_3$CH$_2$OH}).
    \tablewidth{0pt}
    \tablehead{
    \multicolumn{2}{c}{Observed D-ethanol} & \multicolumn{2}{c}{Observed D-glycolaldehyde} & \multicolumn{2}{c}{D-glycol/D-ethanol} \\
    \colhead{Isotopomers} & \colhead{obs. D/H$^{a}$} & \colhead{Isomer} & \colhead{obs. D/H$^{a}$} & \colhead{Observed} & \colhead{Predicted}  
    }
    \startdata
    CH$_3$CH$_2$OD  & 0.05 & CH$_2$ODCHO  & 0.05 & 1.0$\pm$0.8 & 0.90  \\
    CH$_3$CHDOH     & 0.10 & CHDOHCHO     & 0.10 & 1.0$\pm$0.8 & 0.95  \\
    CH$_2$DCH$_2$OH & 0.17 & CH$_2$OHCDO  & 0.05 & 0.30$\pm$0.24 & 0.54\\
    \enddata
\tablecomments{{a) \cite{Jorgensen2018} report an uncertainty of 20\% on the estimates of the ethanol isotopologues column densities, which translates into 40\% on their column density ratio. 
We assumed the same for glycolaldehyde. 
When then propagating to the ratio of the ratios in the penultimate column gives an approximate uncertainty of 80\% on each quoted value.}}
\end{deluxetable*}

Table \ref{tab:astro-comparison} summarises the measured D/H abundance ratios of the three deuterated ethanol isotopomers and the three isotopomers of deuterated glycolaldehyde that would originate from them according to the scheme illustrated in Fig. \ref{fig:reac-scheme}.
Please note that the abundance of a-a and a-s CH$_2$DCH$_2$OH have been added because, even if they can be spectroscopically identified, they are chemically indistinguishable.
In the same table, we report the abundance ratios of the measured and computed (CH$_2$ODCHO/CH$_2$OHCHO)/ (CH$_3$CH$_2$OD/CH$_3$CH$_2$OH), (CHDOHCHO/ CH$_2$OHCHO)/(CH$_3$CHDOH/CH$_3$CH$_2$OH) and (CH$_2$OHCDO/CH$_2$OHCHO)/(CH$_2$DCH$_2$OH/ CH$_3$CH$_2$OH), respectively. 
Please note that by "measured D/H abundance ratios" we mean the ratios of the measured column densities of each D-bearing isotopomer with respect to the H-bearing species. 
Likewise, the computed abundance ratios are obtained considering the rate coefficients of all reactions leading to each deuterated glycolaldehyde isotopomer from the relevant deuterated ethanol one (scheme in Fig. \ref{fig:reac-scheme}).
Also note that the computed ratios actually do not depend on the temperature in the 10--300 K range.

While the measured (CH$_2$ODCHO/CH$_2$OHCHO)/ (CH$_3$CH$_2$OD/CH$_3$CH$_2$OH) and (CHDOHCHO/ CH$_2$OHCHO)/(CH$_3$CHDOH/CH$_3$CH$_2$OH) are about 1, the (CHDOHCHO/ CH$_2$OHCHO)/ (CH$_3$CHDOH/CH$_3$CH$_2$OH) is a factor 3 lower.
The computed abundance ratios follow exactly this trend, with the first two ratios being also around unity, and the last one a factor of two lower.
When considering the error bars associated with the observations (around 20\% for each column density, which translates into a 40\% for two column densities ratios and, finally, to 80\% for the ratio of two ratios) the agreement between the measured and computed abundance ratios is very good.
We notice that \cite{Skouteris2018} already showed that the measured gaseous abundances of ethanol and glycolaldehyde observed in low-mass warm objects are also consistent with the gas-phase formation of glycolaldehyde from ethanol according the scheme of Fig. \ref{fig:ethanol-tree} (see their Figure 4).
The present results support the suggestion by Skouteris et al (2018) because they show that the observed deuteration of ethanol and glycolaldehyde are compatible with a gas-phase synthesis of glycolaldehyde from ethanol.
However, since at present and to our best knowledge, no similar computations on the molecular deuteration for the grain-surface formation routes of iCOMs are available, we cannot exclude that pathway.
It would be extremely important, if not essential, to have also those computations so as to be able to really understand if the two pathways lead to different molecular deuteration and, consequently, if iCOMs deuteration can be used to discriminate their formation route.

Finally, an important result of our new calculations is that, in gas-phase reactions, the  molecular deuteration of the parent species is not "just" passed to the daughter one, as already found by \cite{Skouteris2017} for formamide.
What more, it is impossible to predict a priori the molecular deuteration of different isomers, only specific calculations can provide the correct values.
For example, for the NH$_2$ + H$_2$CO $\rightarrow$ NH$_2$CHO + H$_2$ reaction, the rate coefficients to form NH$_2$CDO from HDCO is about a factor three smaller than that to form NH$_2$CHO, whereas the formation of CH$_3$OHCDO from CH$_2$DCH$_2$OH is a factor two smaller.
Therefore, specific calculations need to be carried out for the systems where deuterated iCOMs isomers are detected, both to have the deuterium fractionation from gas-phase and grain-surface reactions.

%%%%%%%%%%%%%%%%%%%%%%%%%%%%%%%%%%%%%%%%%%%%%%%%%%%%%%%%%%%%%%%%%%%%%%%%%
%%%%%%%%%%%%%%%%%%%%%%%%%%%%%%%%%%%%%%%%%%%%%%%%%%%%%%%%%%%%%%%%%%%%%%%%%
%%%%%%%%%%%%%%%%%%%%%%%%%%%%%%%%%%%%%%%%%%%%%%%%%%%%%%%%%%%%%%%%%%%%%%%%%
\section{Conclusions} \label{sec:Conclusions}

In this work, we present a theoretical study of the gas-phase formation of the deuterated glycolaldehyde isotopomers using the deuterated ethanol isotopomers as mother species. 
This work is based on the original scheme proposed by \cite{Skouteris2018} (see Fig. \ref{fig:ethanol-tree}).
The list of the reactions between the deuterated isotopomers of ethanol and glycolaldehyde is summarised in Tab. \ref{tab:reaction-list} and their scheme is shown in Fig. \ref{fig:reac-scheme}.
It involves two steps which lead first from ethanol to the  2-hydroxyethyl radical (CH$_2$CH$_2$OH) and then from the latter to glycolaldehyde.
Here we report new computations of the second step, which allow to provide theoretical predictions of the abundance ratios of the different D isotopomers of glycolaldehyde with respect to those of ethanol.

We have found a significant primary KIE (kinetic isotope effect) for the case of the reaction CHDCH$_2$OH + O, while in the cases of the other reactions the KIE is modest.
Starting from CH$_3$CH$_2$OH, CH$_3$CH$_2$OD or CH$_3$CHDOH, similar rates for the formation of the major monodeuterated glycolaldehyde  products are obtained, due to the deuterium atom being only an onlooker in these reaction paths. 
On the contrary, starting from CH$_2$DCH$_2$OH, a competition between hydrogen and deuterium loss leads to a depletion of the corresponding deuterated glycolaldehyde, but with a significant difference in the case of reactions CHDCH$_2$OH + O, where the yield of undeuterated glycolaldehyde is not negligible.

This new study confirms that the molecular deuteration of the parent species is not "just" passed to the daughter one, as also found in the case of the formamide when formed in the the NH$_2$ + H$_2$CO gas-phase reaction \citep{Skouteris2017}). 

Finally, the comparison of the new predicted deuterated glycolaldehyde over ethanol isotopomers with those measured towards the Solar-like protostar IRAS16293 B suggests a gas-phase origin for the glycolaldehyde.
However, lacking similar computations for the glycolaldehyde grain-surface formation route, the latter cannot be excluded.

We conclude emphasizing that the new facilities like ALMA and IRAM/NOEMA are now so sensitive that the deuterated iCOMs isomers can be detected. 
Their observed ratios can potentially be used to discriminate their formation routes if, and only if, computations similar to those reported here are carried out both for the gas-phase and grain-surface reactions.

%%%%%%%%%%%%%%%%%%%%%%%%%%%%%%%%%%%%%%%%%%%%%%%%%%%%%%%%%%%%%%%%%%%%%%%%%
%%%%%%%%%%%%%%%%%%%%%%%%%%%%%%%%%%%%%%%%%%%%%%%%%%%%%%%%%%%%%%%%%%%%%%%%%
%%%%%%%%%%%%%%%%%%%%%%%%%%%%%%%%%%%%%%%%%%%%%%%%%%%%%%%%%%%%%%%%%%%%%%%%%
\acknowledgments
We warmly thank Dr. Joan Enrique-Romero, Luca Mancini, Prof. Fernando Pirani and Prof. Marzio Rosi for very helpful discussions.
This project has received funding from the European Research Council (ERC) under the European Union's Horizon 2020 research and innovation programme, for the Project “The Dawn of Organic Chemistry” (DOC), grant agreement No 741002. 
Most of the computations presented in this paper were performed using the GRICAD infrastructure (https://gricad.univ-grenoble-alpes.fr), which is partly supported by the Equip@Meso project (reference ANR-10-EQPX-29-01) of the programme Investissements d'Avenir supervised by the Agence Nationale pour la Recherche
NB and DS thank the Italian Space Agency for co-funding the Life in Space Project (ASI N. 2019-3-U.O).

%%%%%%%%%%%%%%%%%%%%%%%%%%%%%%%%%%%%%%%%%%%%%%%%%%%%%%%%%%%%%%%%%%%%%%%%%
%%%%%%%%%%%%%%%%%%%%%%%%%%%%%%%%%%%%%%%%%%%%%%%%%%%%%%%%%%%%%%%%%%%%%%%%%
%%%%%%%%%%%%%%%%%%%%%%%%%%%%%%%%%%%%%%%%%%%%%%%%%%%%%%%%%%%%%%%%%%%%%%%%%
\bibliography{D-glyco-Vazart}{}
\bibliographystyle{aasjournal}

%%%%%%%%%%%%%%%%%%%%%%%%%%%%%%%%%%%%%%%%%%%%%%%%%%%%%%%%%%%%%%%%%%%%%%%%%
%%%%%%%%%%%%%%%%%%%%%%%%%%%%%%%%%%%%%%%%%%%%%%%%%%%%%%%%%%%%%%%%%%%%%%%%%
%%%%%%%%%%%%%%%%%%%%%%%%%%%%%%%%%%%%%%%%%%%%%%%%%%%%%%%%%%%%%%%%%%%%%%%%%
\clearpage
\appendix

\section{ENERGIES}

\begin{longrotatetable}
\begin{deluxetable*}{lcccccccc}
\tablecaption{Involved energies Part 1, in Hartrees.}
\tablewidth{700pt}
\tabletypesize{\scriptsize}
\tablehead{
\colhead{} & \multicolumn{2}{c}{\textbf{Electronic energies}} & 
\multicolumn{3}{c}{\textbf{undeuterated}} & \multicolumn{3}{c}{\textbf{D1}} \\ 
\colhead{} & \colhead{CCSD(T)/} & \colhead{B2PLYPD3/} & \colhead{ZPE B2} & \colhead{ZPE-corr} & \colhead{ZPE-corr} & \colhead{ZPE B2} & \colhead{ZPE-corr} & \colhead{ZPE-corr} \\
\colhead{} & \colhead{aug-cc-pVTZ} & \colhead{aug-cc-pVTZ} & \colhead{} & \colhead{B2PLYPD3} & \colhead{B2//CCSD(T)} & \colhead{} & \colhead{B2PLYPD3} & \colhead{B2//CCSD(T)}
} 
\startdata
\textbf{CH$_2$CH$_2$OH} & -154.13636 & -154.3101518 & 0.065377849 & -154.244774 & -154.0709822 & 0.0620758 & -154.2480760 & -154.0742842 \\
\textbf{O} & -74.97895233 & -75.05099966 & 0 & \nodata & \nodata & \nodata & \nodata & \nodata \\
\textbf{RI1} & -229.2627622 & -229.511562 & 0.071620018 & -229.439942 & -229.1911422 & 0.0682610 & -229.4433010 & -229.1945012 \\
\textbf{RI2} & -229.2596686 & -229.5087096 & 0.070337589 & -229.438372 & -229.189331 & 0.0670636 & -229.4416460 & -229.1926050 \\
\textbf{TS1} & -229.2555932 & -229.5047613 & 0.070431348 & -229.43433 & -229.1851619 & 0.0671333 & -229.4376280 & -229.1884599 \\
\textbf{RI3} & -229.280304 & -229.5300827 & 0.072307655 & -229.457775 & -229.2079963 & 0.0689597 & -229.4611230 & -229.2113443 \\
\textbf{RI4} & -229.2752895 & -229.5252501 & 0.071691112 & -229.453559 & -229.2035984 & 0.0683771 & -229.456873 & -229.2069124  \\
\textbf{TS10} & -229.2714876 & -229.5214045 & 0.070910492 & -229.450494 & -229.2005771 & 0.0676605 & -229.453744 & -229.2038271 \\
\textbf{cis-glycol.} & -228.7283987 & -228.9792354 & 0.061282384 & -228.917953 & -228.6671163 & 0.0579184 & -228.9213170 & -228.6704803 \\
\textbf{trans-glycol.} & -228.7230381 & -228.9737467 & 0.060606706 & -228.91314 & -228.6624314 & 0.0573357 & -228.9164110 & -228.6657024  \\
\textbf{TS4} & -228.7157718 & -228.9662472 & 0.060708204 & -228.905539 & -228.6550636 & 0.0574292 & -228.9088180 & -228.6583426 \\
\textbf{TS2} & -229.218992 & -229.4698015 & 0.06338652 & -229.406415 & -229.1556054 & 0.0600095 & -229.4097920 & -229.1589824 \\
\textbf{TS3} & -229.2147955 & -229.4655661 & 0.062653149 & -229.402913 & -229.1521423 & 0.0593911 & -229.4061750 & -229.1554043 \\
\textbf{TS6} & -229.2418373 & -229.4951938 & 0.068737806 & -229.426456 & -229.1730995 & 0.0653728 & -229.4298210 & -229.1764645 \\
\textbf{TS5} & -229.2388407 & -229.4922154 & 0.068636375 & -229.423579 & -229.1702044 & 0.0652584 & -229.4269570 & -229.1735824 \\
\textbf{TS13} & -229.1709749 & -229.4211155 & 0.064329474 & -229.356786 & -229.1066454 & 0.0610465 & -229.3600690 & -229.1099284 \\
\textbf{TS14} & -229.1823514 & -229.4323995 & 0.068804454 & -229.363595 & -229.113547 & 0.0658265 & -229.3665730 & -229.1165250 \\
\textbf{TS7} & -229.2134739 & -229.4629519 & 0.066841945 & -229.39611 & -229.1466319 & 0.0634669 & -229.3994850 & -229.1500069 \\
\textbf{TS8} & -229.2097829 & -229.4594576 & 0.066667631 & -229.39279 & -229.1431152 & 0.0633656 & -229.3960920 & -229.1464172 \\
\textbf{TS9} & -229.2156562 & -229.4659675 & 0.067715499 & -229.398252 & -229.1479407 & 0.0643525 & -229.401615 & -229.1513037 \\
\textbf{TS15} & -229.1728045 & -229.4255038 & 0.063381787 & -229.362122 & -229.1094227 & 0.0600308 & -229.3654730 & -229.1127737 \\
\textbf{TS11} & -229.2141893 & -229.4659948 & 0.063609818 & -229.402385 & -229.1505795 & 0.0603338 & -229.4056610 & -229.1538555 \\
\textbf{TS12} & -229.2062338 & -229.4579936 & 0.062820588 & -229.395173 & -229.1434132 & 0.0595416 & -229.3984520 & -229.1466922 \\
\textbf{ep-CH$_2$OCHOH} & -228.6992322 & -228.9483692 & 0.062141192 & -228.886228 & -228.637091 & 0.0588422 & -228.8895270 & -228.6403900 \\
\textbf{E-ethen-1,2-diol} & -228.7083621 & -228.9595503 & 0.061060253 & -228.89849 & -228.6473018 & 0.0577773 & -228.9017730 & -228.6505848 \\
\textbf{Z-ethen-1,2-diol} & -228.7159011 & -228.9669892 & 0.061556193 & -228.905433 & -228.6543449 & 0.0581452 & -228.9088440 & -228.6577559 \\
\textbf{H} & -0.499821176 & -0.498668238 & 0 & \nodata & \nodata & \nodata & \nodata & \nodata \\
\textbf{ethylene oxide} & -153.5565781 & -153.7306824 & 0.057680359 & -153.673002 & -153.4988978 & 0.0543794 & -153.6763030 & -153.5021988 \\
\textbf{OH} & -75.64558477 & -75.7193689 & 0.008511898 & -75.710857 & -75.63707287 & 0.0061969 & -75.7131720 & -75.6393879 \\
\textbf{formaldehyde} & -114.34288 & -114.4688577 & 0.026675717 & -114.442182 & -114.3162043 & 0.0238247 & -114.4450330 & -114.3190553 \\
\textbf{H$_2$COH} & -114.8988513 & -115.0245879 & 0.037422909 & -114.987165 & -114.8614284 & 0.0341269 & -114.9904610 & -114.8647244 \\
\enddata
\end{deluxetable*}
\end{longrotatetable}

\begin{longrotatetable}
\begin{deluxetable*}{lccccccccc}
\tablecaption{Involved energies Part 2, in Hartrees.}
\tablewidth{700pt}
\tabletypesize{\scriptsize}
\tablehead{
\colhead{} & \multicolumn{3}{c}{\textbf{D2}} & 
\multicolumn{3}{c}{\textbf{D3}} & \multicolumn{3}{c}{\textbf{D4}} \\ 
\colhead{} & \colhead{ZPE B2} & \colhead{ZPE-corr} & \colhead{ZPE-corr} & \colhead{ZPE B2} & \colhead{ZPE-corr} & \colhead{ZPE-corr} & \colhead{ZPE B2} & \colhead{ZPE-corr} & \colhead{ZPE-corr} \\
\colhead{} & \colhead{} & \colhead{B2PLYPD3} & \colhead{B2//CCSD(T)} & \colhead{} & \colhead{B2PLYPD3} & \colhead{B2//CCSD(T)} & \colhead{} & \colhead{B2PLYPD3} & \colhead{B2//CCSD(T)}
} 
\startdata
\textbf{CH$_2$CH$_2$OH} & 0.0620798 & -154.2480720 & -154.0742802 & 0.0620478 & -154.2481040 & -154.0743122 & 0.0625128 & -154.2476390 & -154.0738472 \\
\textbf{O} & \nodata & \nodata & \nodata & \nodata & \nodata & \nodata & \nodata & \nodata & \nodata \\
\textbf{RI1} & 0.0682290 & -229.4433330 & -229.1945332 & 0.0682630 & -229.4432990 & -229.1944992 & 0.0685320 & -229.4430300 & -229.1942302 \\
\textbf{RI2} & 0.0669806 & -229.4417290 & -229.1926880 & \nodata & \nodata & \nodata & 0.0672716 & -229.4414380 & -229.1923970 \\
\textbf{TS1} & 0.0670293 & -229.4377320 & -229.1885639 & 0.0670753 & -229.4376860 & -229.1885179 & 0.0674133 & -229.4373480 & -229.1881799 \\
\textbf{RI3} & 0.0689817 & -229.4611010 & -229.2113223 & 0.0689457 & -229.4611370 & -229.2113583 & 0.0692487 & -229.4608340 & -229.2110553 \\
\textbf{RI4} & 0.0683701 & -229.4568800 & -229.2069194 & 0.0683451 & -229.4569050 & -229.2069444 & 0.0686521 & -229.4565980 & -229.2066374  \\
\textbf{TS10} & 0.0675495 & -229.4538550 & -229.2039381 & 0.0676015 & -229.4538030 & -229.2038861 & 0.0679515 & -229.4534530 & -229.2035361 \\
\textbf{cis-glycol.} & 0.0580094 & -228.9212260 & -228.6703893 & \nodata & \nodata & \nodata & 0.0581834 & -228.9210520 & -228.6702153 \\
\textbf{trans-glycol.} & 0.0573127 & -228.9164340 & -228.6657254 & \nodata & \nodata & \nodata & 0.0575827 & -228.9161640 & -228.6654554  \\
\textbf{TS4} & 0.0573502 & -228.9088970 & -228.6584216 & 0.0573392 & -228.9089080 & -228.6584326 & 0.0577352 & -228.9085120 & -228.6580366 \\
\textbf{TS2} & 0.0600335 & -229.4097680 & -229.1589584 & 0.0601215 & -229.4096800 & -229.1588704 & 0.0602595 & -229.4095420 & -229.1587324 \\
\textbf{TS3} & 0.0593101 & -229.4062560 & -229.1554853 & 0.0593541 & -229.4062120 & -229.1554413 & 0.0595801 & -229.4059860 & -229.1552153 \\
\textbf{TS6} & 0.0655148 & -229.4296790 & -229.1763225 & 0.0655518 & -229.4296420 & -229.1762855 & 0.0656418 & -229.4295520 & -229.1761955 \\
\textbf{TS5} & 0.0653994 & -229.4268160 & -229.1734414 & 0.0654434 & -229.4267720 & -229.1733974 & 0.0655344 & -229.4266810 & -229.1733064 \\
\textbf{TS13} & 0.0609195 & -229.3601960 & -229.1100554 & 0.0632915 & -229.3578240 & -229.1076834 & 0.0610185 & -229.3600970 & -229.1099564 \\
\textbf{TS14} & 0.0653815 & -229.3670180 & -229.1169700 & 0.0654045 & -229.3669950 & -229.1169470 & 0.0654905 & -229.3669090 & -229.1168610 \\
\textbf{TS7} & 0.0634609 & -229.3994910 & -229.1500129 & 0.0635699 & -229.3993820 & -229.1499039 & 0.0637089 & -229.3992430 & -229.1497649 \\
\textbf{TS8} & 0.0633366 & -229.3961210 & -229.1464462 & 0.0633176 & -229.3961400 & -229.1464652 & 0.0635386 & -229.3959190 & -229.1462442 \\
\textbf{TS9} & 0.0644655 & -229.4015020 & -229.1511907 & 0.0657745 & -229.4001930 & -229.1498817 & 0.0643685 & -229.4015990 & -229.1512877 \\
\textbf{TS15} & 0.0600088 & -229.3654950 & -229.1127957 & 0.0600808 & -229.3654230 & -229.1127237 & 0.0600658 & -229.3654380 & -229.1127387 \\
\textbf{TS11} & 0.0603418 & -229.4056530 & -229.1538475 & 0.0631358 & -229.4028590 & -229.1510535 & 0.0603118 & -229.4056830 & -229.1538775 \\
\textbf{TS12} & 0.0596056 & -229.3983880 & -229.1466282 & 0.0624006 & -229.3955930 & -229.1438332 & 0.0595566 & -229.3984370 & -229.1466772 \\
\textbf{ep-CH$_2$OCHOH} & 0.0586722 & -228.8896970 & -228.6405600 & 0.0588572 & -228.8895120 & -228.6403750 & 0.0588512 & -228.8895180 & -228.6403810 \\
\textbf{E-ethen-1,2-diol} & 0.0578453 & -228.9017050 & -228.6505168 & \nodata & \nodata & \nodata & \nodata & \nodata & \nodata \\
\textbf{Z-ethen-1,2-diol} & 0.0582662 & -228.9087230 & -228.6576349 & 0.0583272 & -228.9086620 & -228.6575739 & 0.0703752 & -228.8966140 & -228.6455259 \\
\textbf{H} & \nodata & \nodata & \nodata & \nodata & \nodata & \nodata & \nodata & \nodata \\
\textbf{ethylene oxide} & \nodata & \nodata & \nodata & \nodata & \nodata & \nodata & \nodata & \nodata \\
\textbf{OH} & \nodata & \nodata & \nodata & \nodata & \nodata & \nodata & \nodata & \nodata \\
\textbf{formaldehyde} & \nodata & \nodata & \nodata & \nodata & \nodata & \nodata & \nodata & \nodata \\
\textbf{H$_2$COH} & 0.0345569 & -114.9900310 & -114.8642944 & 0.0344919 & -114.9900960 & -114.8643594 & \nodata & \nodata & \nodata \\
\enddata
\end{deluxetable*}
\end{longrotatetable}

\begin{deluxetable}{lccc}
\tablecaption{Involved energies Part 3, in Hartrees.}
\tablewidth{700pt}
\tabletypesize{\scriptsize}
\tablehead{
\colhead{} & \multicolumn{3}{c}{\textbf{D5}} \\ 
\colhead{} & \colhead{ZPE B2} & \colhead{ZPE-corr} & \colhead{ZPE-corr} \\
\colhead{} & \colhead{} & \colhead{B2PLYPD3} & \colhead{B2//CCSD(T)}
} 
\startdata
\textbf{CH$_2$CH$_2$OH} & 0.0625548 & -154.2475970 & -154.0738052 \\
\textbf{O} & \nodata & \nodata & \nodata \\
\textbf{RI1} & 0.0683890 & -229.4431730 & -229.1943732 \\
\textbf{RI2} & \nodata & \nodata & \nodata  \\
\textbf{TS1} & 0.0673133 & -229.4374480 & -229.1882799 \\
\textbf{RI3} & 0.0688927 & -229.4611900	& -229.2114113 \\
\textbf{RI4} & 0.0683401 & -229.4569100 & -229.2069494  \\
\textbf{TS10} & 0.0675745 & -229.4538300 & -229.2039131 \\
\textbf{cis-glycol.} & \nodata & \nodata & \nodata \\
\textbf{trans-glycol.} & \nodata & \nodata & \nodata  \\
\textbf{TS4} & \nodata & \nodata & \nodata \\
\textbf{TS2} & 0.0627545 & -229.4070470 & -229.1562374 \\
\textbf{TS3} & 0.0620581 & -229.4035080 & -229.1527373 \\
\textbf{TS6} & 0.0656298 & -229.4295640 & -229.1762075 \\
\textbf{TS5} & 0.0655574 & -229.4266580 & -229.1732834 \\
\textbf{TS13} & 0.0610165 & -229.3600990 & -229.1099584 \\
\textbf{TS14} & 0.0655005 & -229.3668990 & -229.1168510 \\
\textbf{TS7} & 0.0649069 & -229.3980450 & -229.1485669 \\
\textbf{TS8} & 0.0647136 & -229.3947440 & -229.1450692 \\
\textbf{TS9} & 0.0643795 & -229.4015880 & -229.1512767 \\
\textbf{TS15} & 0.0627688 & -229.3627350 & -229.1100357 \\
\textbf{TS11} & 0.0601708 & -229.4058240 & -229.1540185 \\
\textbf{TS12} & 0.0594846 & -229.3985090 & -229.1467492 \\
\textbf{ep-CH$_2$OCHOH} & \nodata & \nodata & \nodata \\
\textbf{E-ethen-1,2-diol} & \nodata & \nodata & \nodata \\
\textbf{Z-ethen-1,2-diol} & \nodata & \nodata & \nodata \\
\textbf{H} & \nodata & \nodata & \nodata \\
\textbf{ethylene oxide} & \nodata & \nodata & \nodata \\
\textbf{OH} & \nodata & \nodata & \nodata \\
\textbf{formaldehyde} & \nodata & \nodata & \nodata \\
\textbf{H$_2$COH} & \nodata & \nodata & \nodata \\
\enddata
\end{deluxetable}

\section{GEOMETRIES}

\textbf{CH$_2$CH$_2$OH}

 C           1.232920   -0.271107   -0.008952
 
 H           2.125654    0.101853   -0.485779
 
 H           1.288279   -1.185206    0.562369
 
 C          -0.005441    0.539806    0.032131
 
 H          -0.024020    1.254202   -0.796671
 
 H          -0.062224    1.126275    0.954803
 
 O          -1.192797   -0.256587    0.041870
 
 H          -1.150186   -0.856623   -0.708756

\textbf{RI1}

 C       -0.665516    0.600616   -0.243231

 H       -0.680218    0.705686   -1.332409

 H       -1.141144    1.479384    0.188953

 O       -1.413347   -0.525848    0.171215

 H       -0.917795   -1.308601   -0.094782

 C        0.780402    0.508725    0.231273

 H        1.370941    1.396615   -0.029110

 H        0.790069    0.440096    1.334495

 O        1.399450   -0.645306   -0.170639

\textbf{RI2}

 C        0.000000    0.740070    0.000000

 H       -0.485785    1.154026    0.886696

 H       -0.485785    1.154026   -0.886696

 O        1.399587    1.005160    0.000000

 H        1.540173    1.954868    0.000000

 C       -0.153530   -0.771585    0.000000

 H        0.361246   -1.216022   -0.866941

 H        0.361246   -1.216022    0.866941

 O       -1.445826   -1.210383    0.000000

\textbf{TS1}

 C      -0.571160    0.559832    0.050761

 H      -0.372345    1.120069   -0.864881

 H      -0.822570    1.270282    0.834520

 O      -1.715787   -0.276449   -0.081277

 H      -1.639962   -0.778788   -0.897329

 C       0.698158   -0.232201    0.409719

 H       1.040491   -0.005070    1.434793

 H       0.476099   -1.310619    0.443693

 O       1.785324   -0.006259   -0.382932

\textbf{RI3}

 C        -0.673146    0.632927    0.213735

 H        -0.725842    0.737681    1.302874

 H        -1.187125    1.486600   -0.231125

 C         0.726539    0.567617   -0.269325

 H         1.314403    1.450850   -0.461503

 O        -1.362924   -0.601524   -0.045648

 H        -1.404824   -0.716082   -1.001428

 O         1.467182   -0.515066    0.098140

 H         0.848973   -1.229596    0.304778

\textbf{RI4}

 C      -0.583004    0.495629    0.154325

 H      -0.721823    0.973031    1.128737

 H      -0.409785    1.290883   -0.576895

 C       0.580983   -0.424801    0.141213

 H       0.607477   -1.298354    0.780213

 O      -1.807157   -0.194283   -0.090639

 H      -1.692297   -0.711086   -0.894439

 O       1.774329    0.166861   -0.179454

 H       2.491176   -0.460065   -0.050096

\textbf{TS10}

 C       -0.587440   -0.493297    0.194036

 H       -0.369097   -1.299872   -0.506876

 H       -0.827591   -0.958706    1.155437

 C        0.608516    0.390352    0.287982

 H        0.500681    1.465559    0.244071

 O       -1.738098    0.191798   -0.309655

 H       -2.150157    0.670954    0.413262

 O        1.768097   -0.159708   -0.198370

 H        2.479718    0.483011   -0.133811

\textbf{cis-glycol.}

 C         0.826101    0.488051   -0.000000

 H         1.420428    1.416543    0.000002

 O         1.354025   -0.601632   -0.000004

 C        -0.667402    0.649429    0.000002

 H        -0.940852    1.244818    0.879492

 H        -0.940851    1.244816   -0.879490

 O        -1.334428   -0.581691    0.000003

 H        -0.647690   -1.264467    0.000002

\textbf{trans-glycol.}

 C       0.261298   -0.762205    0.000000

 H      -0.647846   -1.389527    0.000000

 O       1.370923   -1.236843    0.000000

 C       0.000000    0.726038    0.000000

 H       0.477464    1.154604    0.885555

 H       0.477464    1.154604   -0.885555

 O      -1.405564    0.916839    0.000000

 H      -1.597739    1.857353    0.000000

\textbf{TS4}

 C         -0.812122    0.368151   -0.259305

 H         -1.150065    1.051941   -1.062078

 O         -1.544305   -0.495696    0.156713

 C          0.609541    0.573067    0.256378

 H          1.019851    1.505268   -0.140520

 H          0.573258    0.635997    1.342221

 O          1.444623   -0.536066   -0.047054

 H          1.569892   -0.586418   -0.999332

\textbf{TS2}

 C       -0.688803    0.636953   -0.136554

 H       -0.873974    1.005642   -1.156596

 H       -1.032802    1.409301    0.551869

 O       -1.373116   -0.559861    0.109977

 H       -0.723000   -1.270556    0.009685

 C        0.811459    0.453501   -0.037354

 H        1.426434    1.359535   -0.171765

 H        0.963126    0.747216    1.675734

 O        1.311151   -0.664372   -0.093161

\textbf{TS3}

 C        0.530279   -0.547980    0.048479

 H        0.545950   -1.303446   -0.744375

 H        0.449612   -1.057604    1.009549

 O        1.657176    0.310567   -0.025786

 H        2.459270   -0.214772    0.022016

 C       -0.710651    0.291448   -0.182338

 H       -0.409780    1.410898    1.153774

 H       -0.538245    1.206897   -0.779232

 O       -1.835248   -0.123414    0.043464

\textbf{TS5}

 C        -0.774386    0.584499    0.017971

 H        -0.579570    1.100826   -0.912598

 H        -0.617988    1.125411    0.937882

 O        -1.851822   -0.237719    0.074655

 H        -2.067464   -0.565388   -0.804312

 C         1.001170   -0.550592    0.059957

 H         0.732556   -0.963553    1.044888

 H         0.659603   -1.157528   -0.796742

 O         1.915841    0.269817   -0.066741

\textbf{TS6}

 C      -0.916588    0.613010   -0.260055

 H      -0.713418    0.673133   -1.322515

 H      -1.312617    1.488241    0.235862

 O      -1.430244   -0.535764    0.217880

 H      -0.926507   -1.275033   -0.153088

 C       1.095015    0.473522    0.265749

 H       1.396154    1.401634   -0.245349

 H       0.850638    0.588877    1.333191

 O       1.384642   -0.638741   -0.203163

\textbf{TS13}

 C         0.290507    0.236734    0.426816

 H         0.328353    0.131863    1.501216

 H         0.622409    1.781508    0.648934

 O         1.406011   -0.055979   -0.308370

 H         1.739609   -0.916701   -0.035033

 C        -0.980476    0.493968   -0.287966

 H        -1.732150    1.100599    0.212893

 H        -0.855129    0.714721   -1.345648

 O        -0.901421   -0.843547    0.081437

\textbf{TS14}

 C        -0.045627   -0.609930    0.008475

 H        -0.182908   -1.118319    0.943641

 H        -0.170151   -1.160955   -0.903213

 O        -1.838019    0.037287   -0.088497

 H        -1.934111    0.521944    0.744518

 C         0.614408    0.739741   -0.038661

 H         0.512259    1.265008   -0.983789

 H         0.498487    1.368354    0.841693

 O         1.570985   -0.244149    0.030780

\textbf{TS7}

 C        0.728784    0.629676    0.134856

 H        0.850615    0.919019    1.189552

 H        1.213079    1.391987   -0.473265

 O        1.369610   -0.601870   -0.123812

 H        0.789235   -1.295584    0.212010

 C       -0.737342    0.582537   -0.163385

 H       -1.312513    1.505308   -0.123083

 H       -1.183254   -0.282076   -0.982745

 O       -1.407837   -0.587122    0.167400

\textbf{TS8}

 C          0.564374    0.516431   -0.049784

 H          0.451828    1.203331    0.791760

 H         0.592263    1.118908   -0.961026

 O          1.796000   -0.189328   -0.010475

 H          1.853187   -0.661848    0.824899

 C         -0.616696   -0.405575   -0.122698

 H         -0.530144   -1.283838   -0.761200

 H         -1.370792   -0.587540    0.876578

 O         -1.881301    0.132559    0.043461

\textbf{TS9}

 C         0.455914   -0.077084    0.466631

 H        -0.328392   -1.099419    0.203930

 H         0.625977   -0.111446    1.541411

 O         1.580239    0.014524   -0.324934

 H         2.225897   -0.641948   -0.045717

 C        -0.737703    0.680711   -0.014300

 H        -1.214195    1.331601    0.716317

 H        -0.559428    1.206588   -0.951607

 O        -1.462630   -0.552916   -0.197356

\textbf{TS15}

 C        -0.751784    0.738214   -0.143326

 H        -1.403173    1.373093    0.451583

 H        -0.654444    1.081041   -1.172631

 C         0.412592    0.148272    0.480287

 H         0.510380    0.083706    1.550092

 O        -0.986187   -0.667875    0.002001

 H        -2.309370   -1.079459   -0.401618

 O         1.544763   -0.171240   -0.168288

 H         1.423153   -0.064367   -1.118893

\textbf{TS11}

 C       0.625410    0.589386   -0.133367

 H       1.190020    1.473197   -0.384238

 H       0.970127    1.044966    1.808993

 C      -0.715361    0.590378   -0.108089

 H      -1.291331    1.494529   -0.212596

 O       1.272148   -0.634655   -0.132610

 H       2.027377   -0.592390    0.462430

 O      -1.460171   -0.519135    0.084712

 H      -0.852301   -1.268569    0.157323

\textbf{TS12}

 C   -0.517063    0.326741   -0.166063

 H   -0.582308    1.419913    1.561885

 H   -0.439956    1.292068   -0.653798

 C    0.567843   -0.404213    0.122955

 H    0.487488   -1.396628    0.537784

 O   -1.766511   -0.255189   -0.111547

 H   -2.357360    0.325392    0.378451

 O    1.854984   -0.003236   -0.035098

 H    1.879675    0.891481   -0.392520

\textbf{ep-CH$_2$OCHOH}

 C      -0.210676    0.043521    0.466750

 H      -0.312304    0.297001    1.513076

 O      -1.460306   -0.118526   -0.091183

 H      -1.361579   -0.247236   -1.041018

 C       0.991370   -0.616439   -0.032043

 H       1.822431   -0.834787    0.626178

 H       0.919052   -1.230789   -0.921369

 O       0.741336    0.800191   -0.256955

\textbf{E-ethen-1,2-diol}

 C        0.531710   -0.400037    0.000001

 H        0.451489   -1.477582    0.000012

 C       -0.531710    0.400037    0.000002

 H       -0.451489    1.477582   -0.000008

 O        1.844519    0.007939   -0.000018

 H        1.883216    0.969720   -0.000001

 O       -1.844519   -0.007939    0.000020

 H       -1.883216   -0.969720   -0.000029

\textbf{Z-ethen-1,2-diol}

 C        -0.634594    0.634205   -0.000292

 H        -1.197735    1.553175   -0.018610

 C         0.695478    0.609595    0.004670

 H         1.284099    1.510850   -0.006688

 O        -1.324020   -0.565840   -0.051979

 H        -2.200907   -0.453474    0.320977

 O         1.438966   -0.533081    0.008958

 H         0.829659   -1.281988    0.022227

\textbf{ethylene oxide}

 C       -0.730917   -0.375206   -0.000000

 H       -1.263500   -0.589074   -0.917612

 H       -1.263500   -0.589075    0.917611

 C        0.730917   -0.375206    0.000000

 H        1.263500   -0.589074    0.917612

 H        1.263500   -0.589074   -0.917611

 O        0.000000    0.857347    0.000000

\textbf{OH}

 O         0.000000    0.000000    0.108015

 H         0.000000    0.000000   -0.864116

\textbf{Formaldehyde}

 C         -0.000000    0.000000   -0.530112

 H          0.000000    0.936285   -1.111690

 H         -0.000000   -0.936285   -1.111690

 O          0.000000   -0.000000    0.675506

\textbf{H$_2$COH}

 C        -0.684998    0.027451   -0.061564

 H        -1.232946   -0.883715    0.101050

 H        -1.114471    0.992303    0.159997

 O         0.670106   -0.125695    0.020201

 H         1.096556    0.732271   -0.053276
 
 \section{Deuteration loops of reactions (1)-(5)}

\begin{figure}[b]
 \begin{center}
\includegraphics[scale=0.30]{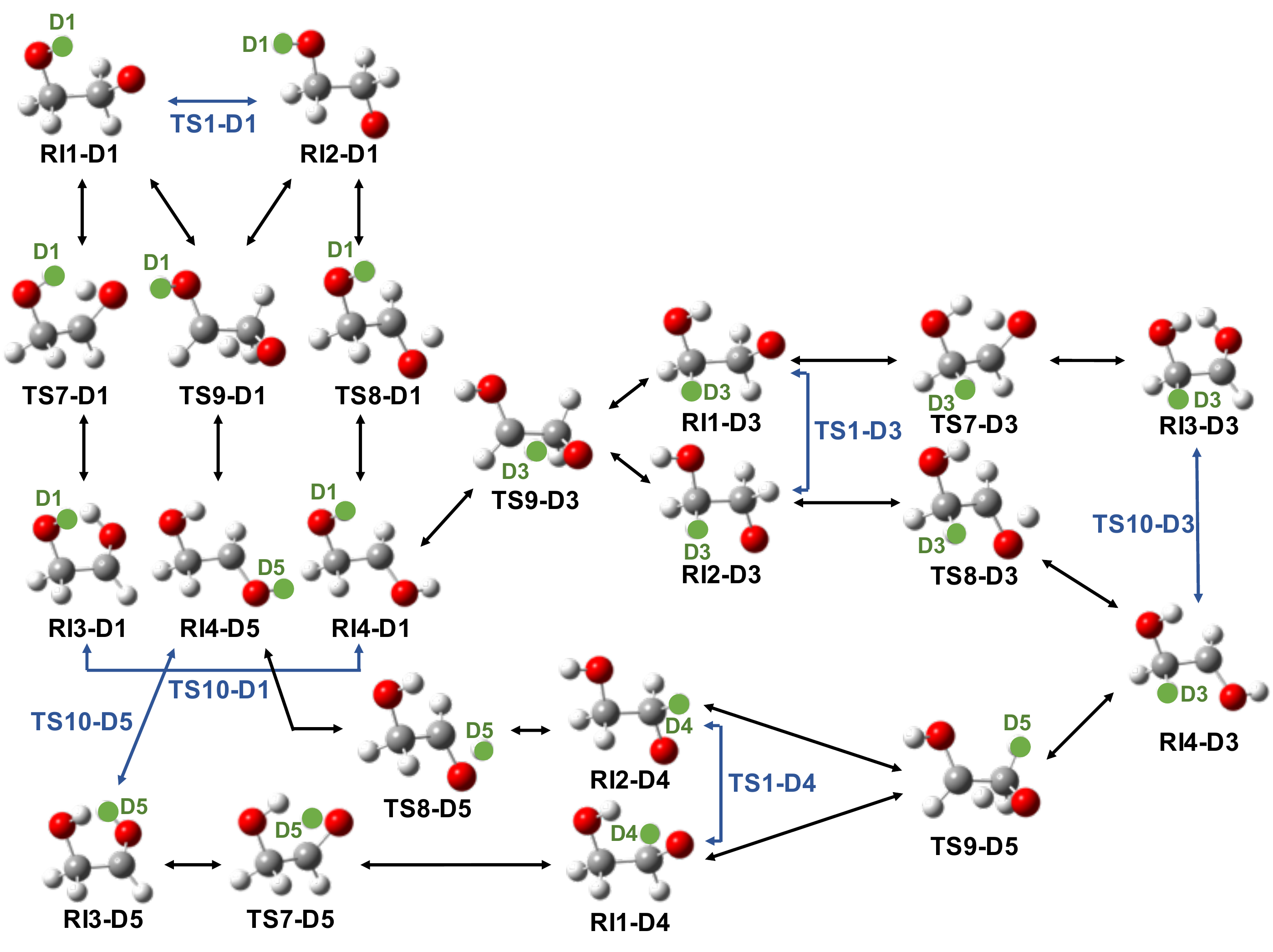}
 \end{center}
\caption{Loop of deuteration regarding reaction (1).}
\label{fig:loop-deu}
\end{figure}

It is noticeable that \textbf{RI1}/\textbf{RI2-D1} leads to \textbf{RI3}/\textbf{RI4-D1} through \textbf{TS7}/\textbf{TS8-D1}, but to \textbf{RI4-D5} through \textbf{TS9-D1}. 
When we look at the backwards step, we have to consider that the \textbf{RI4-D1} leads to \textbf{RI1}/\textbf{RI2-D3} through \textbf{TS9-D3} and that \textbf{RI3}/\textbf{RI4-D5} lead to \textbf{RI1}/\textbf{RI2-D4} though \textbf{TS7}/\textbf{TS8-D5}. 
The last step of the loop is the passage from both \textbf{RI1}/\textbf{RI2-D4} and \textbf{RI1}/\textbf{RI2-D3} to \textbf{RI3}/\textbf{RI4-D3}, through \textbf{TS9-D5} and \textbf{TS7}/\textbf{TS8-D3} respectively. 
For this reaction, it is therefore required to consider the intermediates \textbf{RI1}/\textbf{RI2-D1}/\textbf{D3}/\textbf{D4} and \textbf{RI3}/\textbf{RI4-D1}/\textbf{D3}/\textbf{D5} for the remaining rearrangements and dissociations. 
All of this, keeping in mind that all \textbf{RI1} and \textbf{RI2} are connected by their corresponding \textbf{TS1}, and that all \textbf{RI3} and \textbf{RI4} are connected by their corresponding \textbf{TS10}.

\begin{figure}[h]
 \begin{center}
\includegraphics[scale=0.35]{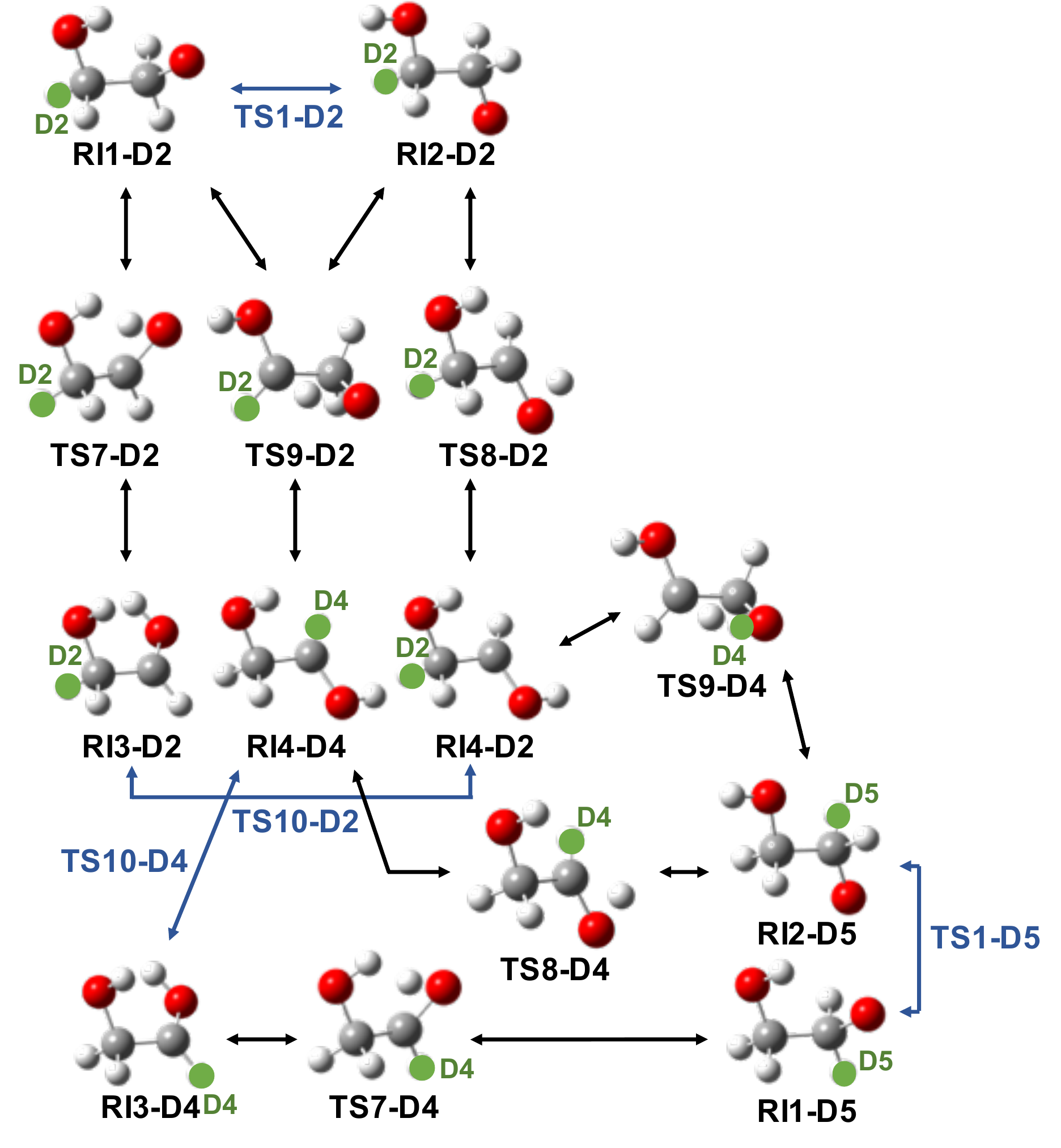}
 \end{center}
\caption{Loop of deuteration regarding reaction (2).}
\label{fig:loop-deu2}
\end{figure}

\begin{figure}[h]
 \begin{center}
\includegraphics[scale=0.27]{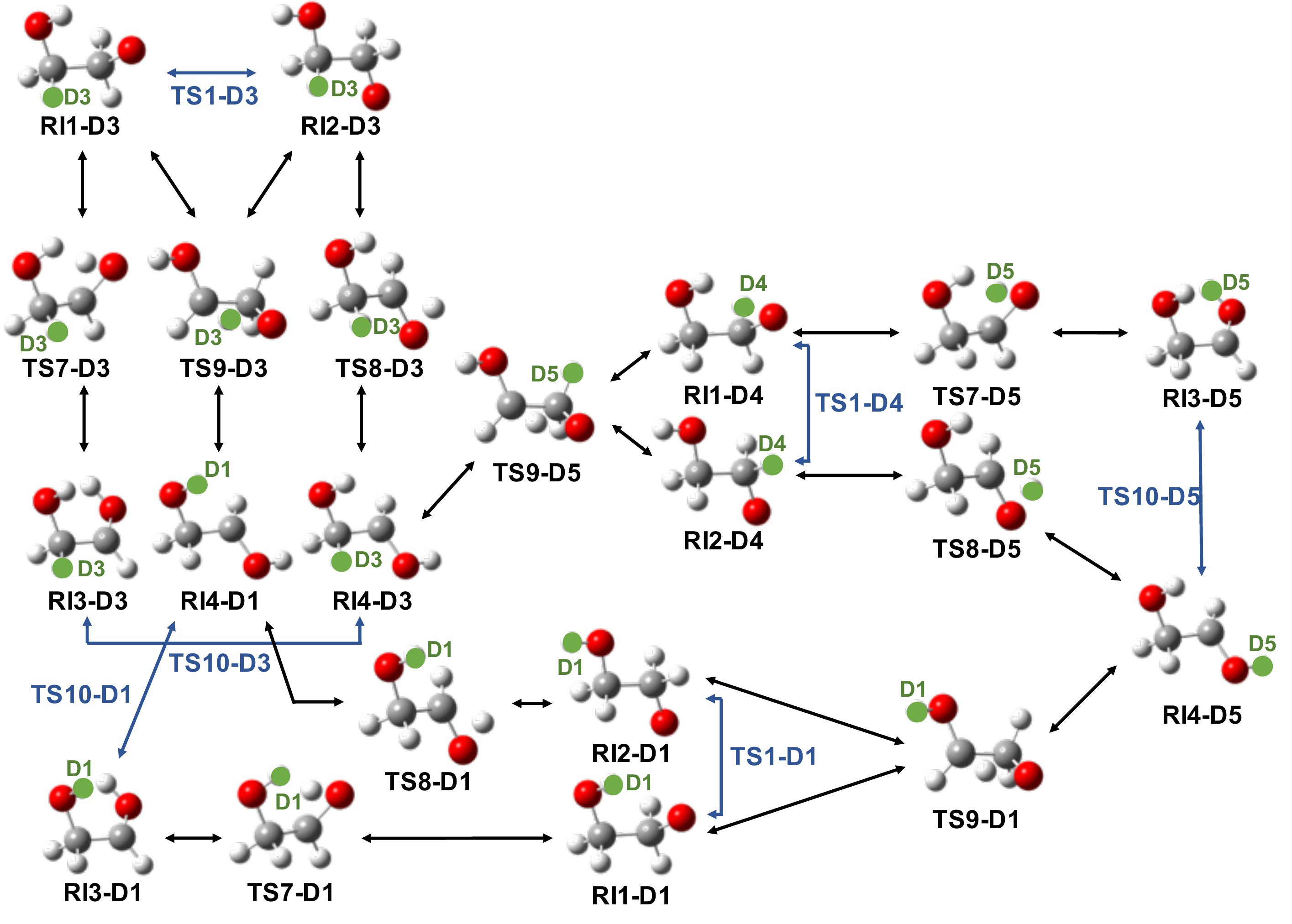}
 \end{center}
\caption{Loop of deuteration regarding reaction (3).}
\label{fig:loop-deu3}
\end{figure}

\begin{figure}[h]
 \begin{center}
\includegraphics[scale=0.30]{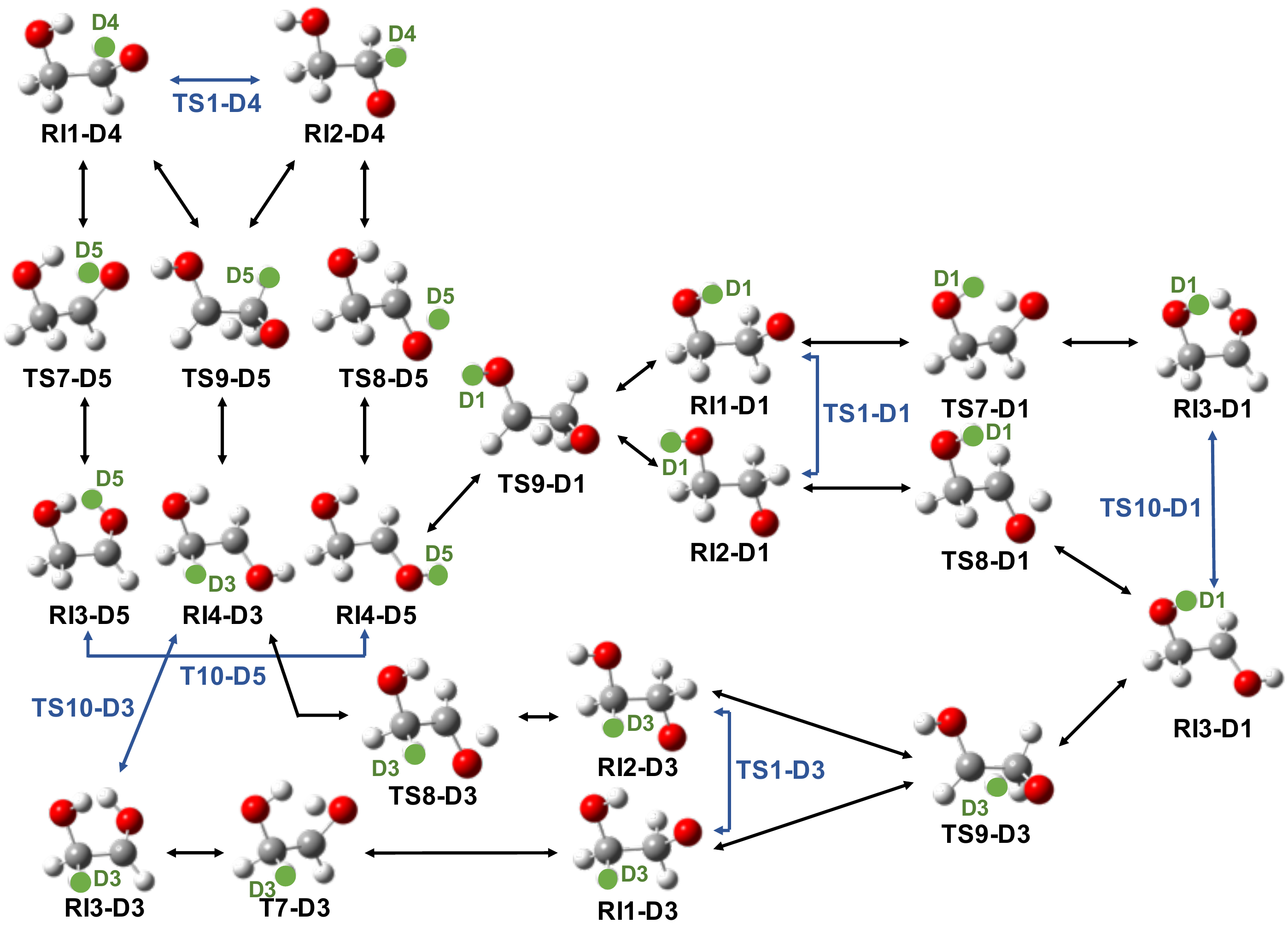}
 \end{center}
\caption{Loop of deuteration regarding reaction (4).}
\label{fig:loop-deu4}
\end{figure}

\begin{figure}[h]
 \begin{center}
\includegraphics[scale=0.35]{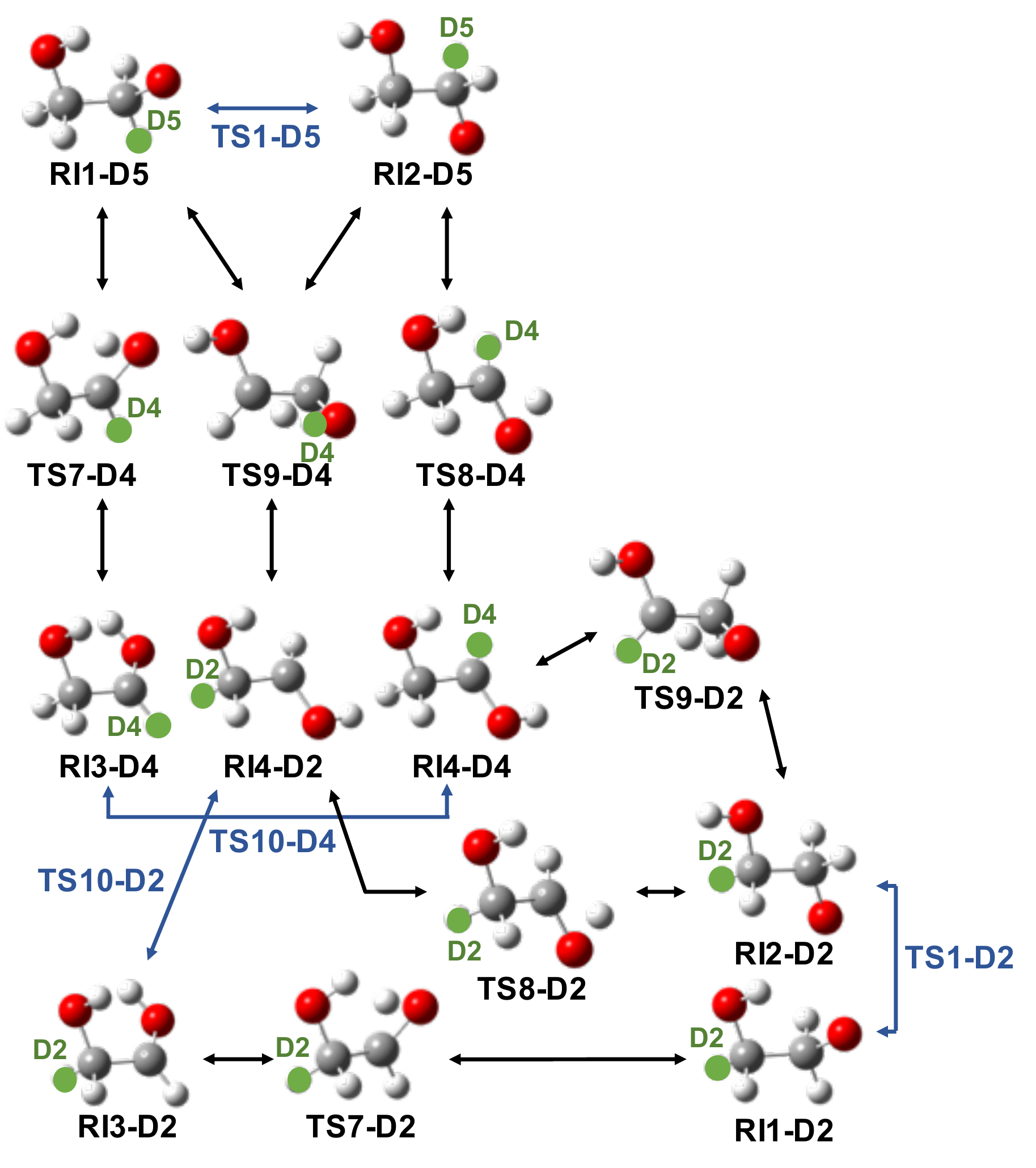}
 \end{center}
\caption{Loop of deuteration regarding reaction (5).}
\label{fig:loop-deu5}
\end{figure}

%% For this sample we use BibTeX plus aasjournals.bst to generate the
%% the bibliography. The sample63.bib file was populated from ADS. To
%% get the citations to show in the compiled file do the following:
%%
%% pdflatex sample63.tex
%% bibtext sample63
%% pdflatex sample63.tex
%% pdflatex sample63.tex

%% This command is needed to show the entire author+affiliation list when
%% the collaboration and author truncation commands are used.  It has to
%% go at the end of the manuscript.
%\allauthors

%% Include this line if you are using the \added, \replaced, \deleted
%% commands to see a summary list of all changes at the end of the article.
%\listofchanges

\end{document}